\documentclass[draftcls,onecolumn,11pt]{IEEEtran}
\usepackage[T1]{fontenc}
\usepackage{rotating,graphics,psfrag,epsfig}
\usepackage{graphicx}

\usepackage{cite}
\usepackage{subfig}
\usepackage{flushend}
\usepackage{xcolor}
\usepackage{amsmath}
\usepackage{amsfonts}
\usepackage{amssymb}
\usepackage[linesnumbered,ruled]{algorithm2e}
\usepackage{array}
\usepackage[printonlyused,withpage]{acronym}

\newcommand{\qed}{\nobreak \ifvmode \relax \else
      \ifdim\lastskip<1.5em \hskip-\lastskip
      \hskip1.5em plus0em minus0.5em \fi \nobreak
      \vrule height0.75em width0.5em depth0.25em\fi}
\usepackage{url}       
\acrodef{DFT}{discrete Fourier transform}
\acrodef{IFFT}{inverse fast Fourier transform}
\acrodef{FFT}{fast Fourier transform}
\acrodef{SNR}{signal-to-noise ratio}
\acrodef{CP}{cyclic prefix}
\acrodef{ISI}{inter symbol interference}
\acrodef{PA}{Power-Amplifier}
\acrodef{i.i.d.}{independent and identically distributed}
\acrodef{mm-wave}{millimeter-wave}
\acrodef{cm-wave}{centimeter-wave}
\acrodef{SMD}{sparsity mask detection}
\acrodef{MIMO}{multiple-input-multiple-output}
\acrodef{SNR}{signal-to-noise ratio}
\acrodef{FIM}{Fisher information matrix}
\acrodef{CRB}{Cram\'{e}r-Rao bound}
\acrodef{LLR}{log-likelihood ratio}
\acrodef{VMF}{Von-Mises-Fisher}
\acrodef{CWNA}{continuous white noise acceleration}
\acrodef{UWB}{ultra-wide bandwidth}
\acrodef{GPS}{global positioning system}
\acrodef{BF}{beamforming}
\acrodef{iid}{independent and identically distributed}
\acrodef{QoS}{quality-of-service}
\acrodef{LOS}{line-of-sight}
\acrodef{ROC}{receiver operating characteristic}
\acrodef{CFAR}{constant false alarm rate}
\acrodef{SD}{support detection}
\acrodef{QCD}{quickest change detection}
\acrodef{TCD}{transient change detection}
\acrodef{FMA}{finite moving average}
\acrodef{WLC}{window-limited CUSUM}
\acrodef{CUSUM}{cumulative sum}
\acrodef{NLOS}{non-line-of-sight}
\acrodef{OLOS}{obstructed-line-of-sight}
\acrodef{RF}{radio-frequency}
\acrodef{EXIP}{extended invariance principle}
\acrodef{DLS}{damped least-squares}
\acrodef{CDF}{cumulative distribution function}
\acrodef{MPCs}{multi-path components}
\acrodef{ML}{maximum likelihood}
\acrodef{MS}{mobile station}
\acrodef{pdf}{probability density function}
\acrodef{WLS}{weighted least squares}
\acrodef{LMA}{Levenberg-Marquardt algorithm}
\acrodef{GNA}{Gauss-Newton algorithm}
\acrodef{BS}{base station}
\acrodef{ADC}{analog-to-digital-converter}
\acrodef{AOA}{angle-of-arrival}
\acrodef{UL-AOA}{uplink AOA}
\acrodef{DL-AOA}{downlink AOA}
\acrodef{UL-AOD}{uplink AOD}
\acrodef{DL-AOD}{downlink AOD}
\acrodef{DOA}{direction-of-arrival}
\acrodef{AOD}{angle-of-departure}
\acrodef{TOA}{time-of-arrival}
\acrodef{TDOA}{time-difference-of-arrival}
\acrodef{ULA}{uniform linear array}
\acrodef{PSD}{positive semidefinite}
\acrodef{EFIM}{equivalent Fisher information matrix}
\acrodef{FB}{fractional bandwidth}
\acrodef{REB}{rotation error bound}
\acrodef{PEB}{position error bound}
\acrodef{SDL}{sensor delay line}
\acrodef{TDL}{tapped delay line}
\acrodef{OMP}{orthogonal matching pursuit}
\acrodef{DCS-SOMP}{distributed compressed sensing-simultaneous orthogonal matching pursuit}
\acrodef{DCS}{distributed compressed sensing}
\acrodef{SS-UKF}{spherical simplex unscented Kalman filter}
\acrodef{UKF}{unscented Kalman filter}
\acrodef{EKF}{extended Kalman filter}
\acrodef{CS}{compressed sensing}
\acrodef{CoSOMP}{compressive sampling matched pursuit}
\acrodef{SOMP}{simultaneous OMP}
\acrodef{RA-ORMP}{rank-aware order recursive matching pursuit}
\acrodef{G-BPDN}{group basis pursuit denoising}
\acrodef{GCS}{group sparse compressed sensing}
\acrodef{MMV}{multiple measurement vectors}
\acrodef{SMV}{single measurement vector}
\acrodef{ReMBo}{reduce MMV and boost}
\acrodef{MLE}{maximum likelihood estimation}
\acrodef{IQML}{iterative quadratic maximum likelihood}
\acrodef{RMSE}{root-mean-square error}
\acrodef{LS}{least squares}
\acrodef{RSE}{root-square error}
\acrodef{rsCRB}{root-square CRB}
\acrodef{RMS}{root-mean-square}
\acrodef{MMSE}{minimum mean square error}
\acrodef{EM}{expectation maximization}
\acrodef{SAGE}{space-alternating generalized expectation maximization}
\acrodef{OFDM}{orthogonal frequency division multiplexing}

\usepackage{graphicx}
\begin{document}                        
\title{Tracking Position and Orientation through Millimeter Wave Lens MIMO in 5G Systems}
\author{Arash Shahmansoori, Bernard Uguen,~\IEEEmembership{Member,~IEEE}, Giuseppe Destino,~\IEEEmembership{Member,~IEEE}, Gonzalo Seco-Granados,~\IEEEmembership{Senior Member,~IEEE}, and Henk Wymeersch,~\IEEEmembership{Member,~IEEE}
\thanks{
Arash Shahmansoori and Bernard Uguen are with the Institute of Electronics and Telecommunications of Rennes, Universit\'{e} de Rennes 1, 35042 Rennes, France, emails: arash.mansoori65@gmail.com and Bernard.Uguen@univ-rennes1.fr. Gonzalo Seco-Granados is with the Department of Telecommunications and Systems Engineering, Universitat Aut\`{o}noma de Barcelona, 08193 Barcelona, Spain, email: gonzalo.seco@uab.cat. Henk Wymeersch is with the Department of Electrical Engineering, Chalmers University of Technology, 412 96 Gothenburg, Sweden, email: henkw@chalmers.se. Giuseppe Destino is with the center for wireless communications, University of Oulu, 90014 Oulu, Finland, and visiting research fellow at King's College London, email:  giuseppe.destino@oulu.fi. This work was financially supported by M5HESTIA (mmW Multi-user Massive MIMO Hybrid Equipments for Sounding, Transmissions and HW ImplementAtion) project, the EU-H2020 project HIGHTS (High Precision Positioning for Cooperative ITS Applications) under grant nr. MG-3.5a-2014-636537, the VINNOVA COPPLAR project, funded under Strategic Vehicle Research and Innovation grant nr. 2015-04849, FALCON (Fundamental of simulatemous localization and communications) funded by the Academy of Finland, and R\&D Projects of Spanish Ministry of Economy and Competitiveness TEC2017-89925-R. (Corresponding author: Arash Shahmansoori.)}
}
\maketitle
\begin{abstract}
Millimeter wave signals and large antenna arrays are considered enabling technologies for future 5G networks. Despite their benefits for achieving high data rate communications, their potential advantages for tracking of the location of the user terminals are largely undiscovered. In this paper,  we propose a novel support detection-based channel training method for frequency selective \ac{mm-wave} multiple-input-multiple-output system with lens antenna arrays. We show that accurate position and orientation estimation and tracking is possible using signals from a single transmitter with lens antenna arrays. Particularly, the beamspace channel estimation is formulated as two sparse signal recovery problems in the downlink and uplink for the estimation of angle-of-arrival, angle-of-departure, and time-of-arrival. The proposed method offers a higher sparse detection probability compared to the compressed sensing based solutions. Finally, a joint heuristic beamformer design and user position and orientation tracking approach are proposed based on initial estimation of channel parameters obtained in the training phase. 
\end{abstract}
\begin{IEEEkeywords}
5G networks, \ac{mm-wave}, lens arrays, support detection-based channel training, position and orientation tracking, heuristic beamformer design.
\end{IEEEkeywords}
\section{Introduction}
\IEEEPARstart{M}{m-wave} and massive \ac{MIMO} will likely be adopted technologies in fifth generation (5G) communication networks, thanks to a number of favorable properties. Particularly, due to exploiting the carrier frequencies beyond 30 GHz and large available bandwidth, \ac{mm-wave} can provide high data rate. This can be obtained through dense spatial multiplexing with large antennas. \cite{Zhouyue,Rappaport}. Despite the aforementioned properties that are desirable for 5G services, there are a number of challenges regarding \ac{mm-wave} communications. One of the most important challenges is the severe path loss at high carrier frequencies. The loss in \ac{SNR} is compensated through beamforming at the transmitter and/or receiver resulting in highly directional links \cite{Wang,Hur,Tsang}. The design of beamformers requires the knowledge of propagation channel, e.g., user position, scatterer locations, and so on. Since at \ac{mm-wave} frequencies only the \ac{LOS} and a few dominant multipath components contribute to the received signal power, the channel is sparse in the angular domain \cite{BspaceSayeed,widebandbrady}. In other words, diffuse scattering and multiple-bounce reflections are much weaker than \ac{LOS} and single-bounce specular reflection \cite{Martinez-Ingles,Vaughan,mmMAGIC}. Different methods for \ac{mm-wave} channel estimation have been proposed by exploiting sparsity \cite{Marzi,LeeJ,AlkhateebA,ChoiJ,AlkhateebC,HanY,LeeJ2,Ramasamy,BerrakiD} and compressed sensing tools such as \ac{DCS-SOMP} \cite{Duarte}, \ac{CoSOMP} \cite{Duarte2}, and \ac{GCS} \cite{Bolcskei}. In \cite{Marzi}, a method for channel parameter estimation is proposed based on the downlink compressive beacons. A \ac{CS}-based method with redundant dictionary is proposed in \cite{LeeJ}. A method based on hierarchical multi-resolution codebook design for the estimation of single-path and multi-path \ac{mm-wave} channels is proposed in \cite{AlkhateebA}. In \cite{ChoiJ}, multiuser \ac{mm-wave} \ac{MIMO} channels with analog beamformers using a beam selection procedure is proposed. A \ac{CS}-based method with reduced training overhead was considered in \cite{AlkhateebC}. A robust algorithm with single feedback is used in \cite{HanY}. A method based on \ac{CS} tools with angular refinement is applied in \cite{LeeJ2}, and continuous estimation of \ac{mm-wave} channel parameters is proposed in \cite{Ramasamy}. In \cite{BerrakiD}, \ac{CS} tools are used for estimation of power-angle profiles of the \ac{mm-wave} channels with overhead reduction compared to the codebook designs. In the aforementioned papers, a narrow-band \ac{mm-wave} channel is considered. In this paper, we extend the results to the wideband \ac{mm-wave} channel model, i.e., the delays of different paths are also estimated.

In \ac{mm-wave} frequencies, and due to large number of antenna elements, equipping each antenna with a dedicated \ac{RF}-chain is not cost-effective in terms of implementation and power consumption. To reduce the number of \ac{RF}-chains, there are two approaches. In the first approach, named as hybrid analog/digital processing, the digital precoding and combining is performed in the baseband followed by phase shifters in the \ac{RF} band \cite{GaoXx}. However, this approach may suffer from  inefficiency of \ac{PA} at high carrier frequency, non-linearity of \ac{PA}, and heat dissipation and imbalances/non-ideality of the phase shifter responses. In the second approach, a low-cost implementation of \ac{mm-wave} \ac{MIMO} is achieved by using switching circuits together with lens antenna arrays \cite{ZengYy,ZengYy2,BehdadNn,BradySay2,Xinyu_Gau,Yang_Lu}. To gain low-cost implementation with \ac{mm-wave} lens \ac{MIMO}, it is important to find efficient channel estimation methods. In \ac{mm-wave} lens \ac{MIMO} systems, the analog precoders are often in practice restricted to \ac{DFT} vectors \cite{BehdadNn}. Moreover, adopting traditional channel estimation techniques from hybrid analog/digital systems (e.g., the exhaustive and hierarchical search) for \ac{mm-wave} lens \ac{MIMO} systems does not lead to satisfying results. Different methods are proposed for \ac{mm-wave} lens \ac{MIMO} channel estimationin \cite{Hoganjj,Xinyu_Gau,Yang_Lu}. In \cite{Hoganjj}, the \ac{SMD} \ac{mm-wave} lens \ac{MIMO} channel estimation is proposed. In this approach, the beams with larger energy are detected with a beam training procedure between users and the \ac{BS}. Then, the dimension-reduced beamspace channel is estimated through \ac{LS} solutions. The computational complexity and pilot overhead in this method is reduced, however the number of required pilot symbols to scan all the beams is on the order of the number of antenna elements, e.g., 256 antennas. A similar approach based on energy based detection in uplink and downlink was used in \cite{Yang_Lu}. In \cite{Xinyu_Gau}, the \ac{SD}-based estimation with reduced pilot symbols for narrowband \ac{mm-wave} channel is proposed by exploiting the sparsity in the beamspace. The method performs better than the \ac{CS} based approaches as it relies on the power detection. This results in better estimation capabilities at low \ac{SNR}. However, the method is limited to the estimation of \ac{AOA} and beamspace coefficients for the narrowband \ac{mm-wave} lens \ac{MIMO} channel. 

The relative location of transmitter and receiver can be obtained using the estimated \ac{AOA}/\ac{AOD} \cite{USPatent1}. Moreover, the estimated user location provides necessary information for beamformer design. In other words, once the user location is estimated, the \ac{BS} can steer the beam towards the user, either directly using the \ac{LOS} path or through the first-order reflectors. This results in a synergy between localization and communication. The position and orientation estimation was previously explored in \cite{sanchis2002novel,DenSaya,vari2014mmwaves} and in \cite{hu2014esprit,Dardari,savic2015fingerprinting} for \ac{mm-wave} and massive \ac{MIMO} systems, respectively. In \cite{sanchis2002novel}, a beam switching approach was suggested for tracking of \ac{AOA}. A link by link \ac{mm-wave} \ac{AOA}/\ac{AOD} and channel gain tracking was proposed in \cite{VaV}, while a tracking solution for all the links was investigated in \cite{ZhangC}. In \cite{DenSaya}, user localization was formulated as a hypothesis testing problem for a given spatial resolution. Meter-level positioning accuracy was obtained using the received signal strength levels in \cite{vari2014mmwaves}. To speed up initial access between nodes, a location-aided beamforming method was proposed in \cite{NGarcia}. In the massive \ac{MIMO} case, \cite{hu2014esprit} considered the estimation of angles, a direct localization methods by jointly processing the observations at the distributed massive \ac{MIMO} \ac{BS}s was proposed in\cite{NGarcia2}, while joint estimation of \ac{AOA}/\ac{AOD} and \ac{TOA} in the \ac{LOS} with the impact of errors in delays and phase shifters was investigated in\cite{Dardari}, sufficient conditions to obtain a non-singular \ac{FIM} for delay, \ac{AOA}/\ac{AOD} and channel coefficients were derived in\cite{Arash}, and \cite{Arash2x} proposed the corresponding estimators that approach \ac{PEB} and \ac{REB} for average to high \ac{SNR} in the \ac{LOS}, \ac{NLOS}, and \ac{OLOS}   conditions. By linearizing the non-linear equations for \ac{TDOA}, \ac{AOA}, and \ac{AOD}, a hybrid localization approach was proposed in \cite{linhyb}. In \cite{savic2015fingerprinting}, a Gaussian process regressor was applied to solve the positioning by operating on a vector of received signal strengths through fingerprinting. This approach does not harness the geometry of the environment, while it exploits the \ac{NLOS} propagation. Complementarily to the use of \ac{mm-wave} frequencies, approaches for localization using \ac{cm-wave} signals have been recently proposed as well. Joint \ac{TDOA}s and \ac{AOA}s using an \ac{EKF} was proposed in \cite{DBLP:journals/twc/KoivistoCWHTLKV17,DBLP:journals/corr/KoivistoHCKLV16}. In this method, the \ac{MS} has a single antenna and the \ac{BS} employs an antenna array to provide sub-meter accuracy for moving devices. The main assumption is the \ac{LOS} propagation thanks to high density of access nodes.

In this paper, we propose a position and orientation training method with one \ac{BS} followed by the tracking phase. The method considers joint tracking of position and orientation together with beamformer design in the tracking phase initialized by the estimated values from the training phase. The main contributions of the paper are:
\begin{itemize}
\item A novel \ac{SD}-based \ac{mm-wave} channel estimation for frequency selective lens \ac{MIMO} is proposed in the training phase. The main advantage of the proposed method over the traditional adopted \ac{CS}-based (e.g., exhaustive and hierarchical search) \ac{mm-wave} channel estimation approaches for lens \ac{MIMO} is the increased detection probability of the sparse channel support, i.e., improved estimation accuracy. Moreover, the proposed method reduces the required time for training compared to the \ac{CS}-based approaches.
\item A position and orientation tracking method is proposed based on the \ac{EKF} and the heuristic beamformer design in the tracking phase. The proposed method enables tracking of the position and rotation angle of the user with reduced number of required beams within the observation time.
\item Joint heuristic beamformer and position and orientation training-tracking algorithm is proposed. From the simulation results, it is observed that the proposed algorithm provides comparable accuracies to the corresponding values of the \ac{PEB} and \ac{REB} even at low \ac{SNR}.
\end{itemize}

\begin{figure}   
\psfrag{x}{\small $x$}
\psfrag{y}{\small  $y$}
\psfrag{d0}{\small $d_{0}$}
\psfrag{dk1}{\small  $d_{k,1}$}
\psfrag{dk2}{\small $d_{k,2}$}
\psfrag{dk}{\small $d_{k}=d_{k,1}+d_{k,2}$}
\psfrag{sk}{\hspace{-1mm} $\mathbf{s}_{k}$}
\psfrag{tt0}{\small \hspace{-1mm} $\theta_{0}$}
\psfrag{tt1}{\small  \hspace{-4mm} $\theta_{k}$}
\psfrag{tt1b}{\small  \hspace{-8mm} $\pi-(\phi_{k}+\alpha)$}
\psfrag{rr0}{\small  \hspace{-8mm} $\pi-\phi_{0}$}
\psfrag{rr1}{\small \hspace{-8mm} $\pi-\phi_{k}$}
\psfrag{rr1b}{\small \hspace{-8mm} $\pi-(\phi_{k}+\alpha)$}
\psfrag{alphab}{\small  \hspace{-4mm} $\alpha$}
\psfrag{q}{ \hspace{-1mm} $\mathbf{q}$}
\psfrag{qk}{ \hspace{-1mm} $\widetilde{\mathbf{q}}_{k}$}
\psfrag{p}{  \hspace{-4mm} $\mathbf{p}$}
\psfrag{BS}[][c]{BS}
\psfrag{VBS}[][c]{virtual BS}
\psfrag{MS}[][c]{MS}
\centering
\includegraphics[width=0.7\columnwidth]{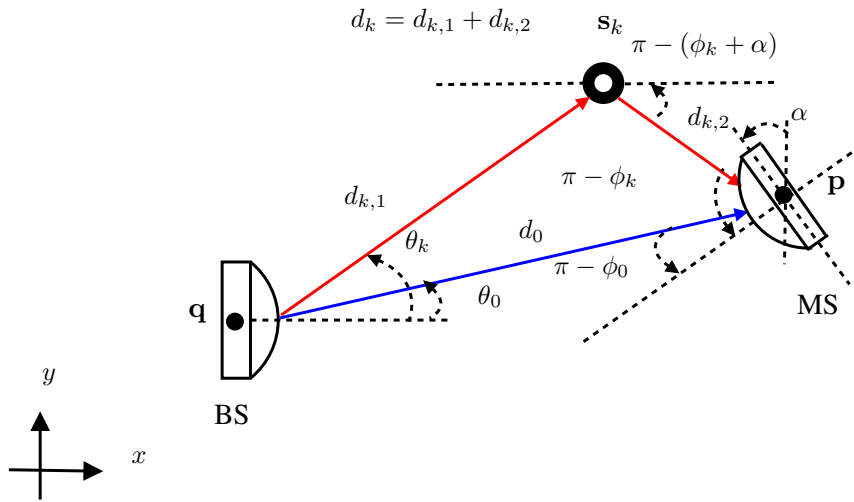}
  \caption{Two dimensional illustration of the \ac{LOS} (blue link) and \ac{NLOS} (red link) based positioning problem. The \ac{BS} location $\mathbf{q}$ and \ac{BS} orientation are known, but arbitrary. The unknown parameters include: the location of the \ac{MS} $\mathbf{p}$, scatterer $\mathbf{s}_{k}$, rotation angle $\alpha$, \ac{DL-AOA}s $\{\phi_{k}\}$, \ac{UL-AOA}s $\{\theta_{k}\}$, the channels between \ac{BS}, \ac{MS}, and scatterers, and the distance between the antenna centers.}
  \label{NLOS_Link}
\end{figure}
\section{System Model}\label{Sys_Mod}
A \ac{MIMO} system with a \ac{BS} equipped with $N_{\mathrm{BS}}$ antennas and a \ac{MS} equipped by $N_{\mathrm{MS}}$ antennas operating at a carrier frequency $f_c$ (corresponding to wavelength $\lambda_c$) and bandwidth $B$ has been considered. 
The \ac{BS} and \ac{MS} locations are denoted by $\mathbf{q}=[q_{x}, q_{y}]^{\mathrm{T}}\in\mathbb{R}^{2}$ and $\mathbf{p}=[p_{x}, p_{y}]^{\mathrm{T}}\in\mathbb{R}^{2}$. The rotation angle of the \ac{MS}'s antenna array is denoted by $\alpha\in[0, 2\pi)$. The location of the \ac{BS} $\mathbf{q}$ is assumed to be known, while $\mathbf{p}$ and $\alpha$ are unknown and need to be estimated. 
\subsection{Transmitter Model}
It is assumed that \ac{OFDM} signals are used for transmission as in \cite{khateeb3}. A \ac{BS} with hybrid analog/digital precoder and lens array communicates with a single \ac{MS} with lens array. The \ac{BS} sequentially transmits $G$ signals. The $g$-th transmission comprises  $M_{\mathrm{BS}}$ simultaneously transmitted symbols $\mathbf{x}^{(g)}[n]=\frac{1}{\sqrt{M_{\mathrm{BS}}}}[x_{1}[n],\ldots,x_{M_{\mathrm{BS}}}[n]]^{\mathrm{T}} \in \mathbb{C}^{M_{\mathrm{BS}}}$ for each subcarrier $n=0,\ldots,N-1$. The precoded symbols are transformed to the time-domain using $N$-point \acf{IFFT}. A \acf{CP} of length $T_{\mathrm{CP}}=DT_{s}$ is added before applying the \ac{RF} precoding where $D$ is the length of \ac{CP} in symbols. The sampling period is defined as $T_{s}=1/B$ and  $T_{\mathrm{CP}}$ exceeds the channel delay spread. The transmitted signal over subcarrier $n$ at the $g$-th transmission can be expressed as $\mathbf{F}^{(g)}_{\mathrm{RF},\mathrm{BS}}\mathbf{F}^{(g)}_{\mathrm{BB},\mathrm{BS}}\mathbf{x}^{(g)}[n]$ where $\mathbf{F}^{(g)}_{\mathrm{RF},\mathrm{BS}}$ is the analog precoder. The term $\mathbf{F}^{(g)}_{\mathrm{BB},\mathrm{BS}}$ denotes the digital baseband precoder to compensate for imperfections of the desired beamformer design in the analog. The total power constraint $\Vert\mathbf{F}^{(g)}_{\mathrm{RF},\mathrm{BS}}\mathbf{F}^{(g)}_{\mathrm{BB},\mathrm{BS}}\Vert_{\mathrm{F}}=1$ is satisfied by the transmit beamformer. 
\subsection{Mm-Wave Lens MIMO Channel Model}
Fig.~\ref{NLOS_Link} shows the position-related parameters of the channel including the location of the $k$-th scatterer $\mathbf{s}_{k}$, $\mathbf{p}$, and $\mathbf{q}$. Also, the corresponding channel parameters include $\phi_{k}$, $\theta_{k}$, and $d_{k}=c\tau_{k}$, denoting the \ac{AOA}, \ac{AOD}, and the path length (with \ac{TOA} $\tau_{k}$ and the speed of light $c$) of the $k$-th path ($k=0$ for the \ac{LOS} path and  $k>0$ the \ac{NLOS}  paths). For the $k$-th NLOS path, we define $d_{k,1}=\Vert\mathbf{s}_{k}-\mathbf{q}\Vert_{2}$ and $d_{k,2}=\Vert\mathbf{p}-\mathbf{s}_{k}\Vert_{2}$. We consider a frequency-independent array response \cite{widebandbrady}. The channel is assumed to be constant during the transmission of $G$ symbols in the downlink. Assuming $K+1$ paths, the $N_{\mathrm{MS}}\times N_{\mathrm{BS}}$ \ac{mm-wave} \ac{MIMO} channel in the downlink is expressed as 
\begin{equation}\label{Channel1}
 \mathbf{H}_{\mathrm{DL}}[n]=\mathbf{A}_{\mathrm{MS}}\mathbf{\Gamma}[n]\mathbf{A}^{\mathrm{H}}_{\mathrm{BS}},
 \end{equation}
where 
 \begin{align}
 \mathbf{A}_{\mathrm{BS}}& =[\mathbf{a}_{\mathrm{BS}}(\theta_{0}),\ldots,\mathbf{a}_{\mathrm{BS}}(\theta_{K})], \\
 \mathbf{A}_{\mathrm{MS}}& =[\mathbf{a}_{\mathrm{MS}}(\phi_{0}),\ldots,\mathbf{a}_{\mathrm{MS}}(\phi_{K})],\\
 \mathbf{\Gamma}[n]& =\mathrm{diag}\left\{\gamma_{n}(\tilde{h}_{0},\tau_{0}),\ldots,\gamma_{n}(\tilde{h}_{K},\tau_{K})\right\},
 \end{align}
 where $\gamma_{n}(\tilde{h}_{k},\tau_{k})$ is defined as $\gamma_{n}(\tilde{h}_{k},\tau_{k})=\tilde{h}_{k}e^{-j2\pi n\tau_{k}/(NT_{s})}$ and $\tilde{h}_{k}=\sqrt{(N_{\mathrm{BS}}N_{\mathrm{MS}})/\rho_{k}}h_{k}$ in which $\rho_{k}$ denotes the path loss. The steering vector is defined as
\begin{equation}\label{steering_vector1}
\mathbf{a}_{\mathrm{BS}}(\theta_{k})=\frac{1}{\sqrt{N_{\mathrm{BS}}}}[
e^{-j \frac{N_{\mathrm{BS}}-1}{2}\frac{2\pi}{\lambda_{c}}d\sin(\theta_{k})},\ldots,e^{j\frac{N_{\mathrm{BS}}-1}{2}\frac{2\pi}{\lambda_{c}}d\sin(\theta_{k})}
]^{\mathrm{T}}.
\end{equation}
The term $\mathbf{a}_{\mathrm{MS}}(\phi_{k})$ is defined similarly by replacing the subscript $\mathrm{BS}$ by $\mathrm{MS}$, and $\theta_{k}$ by $\phi_{k}$. The \ac{mm-wave} lens \ac{MIMO} channel model is obtained using \ac{DFT} matrices $\mathbf{U}_{\mathrm{BS}}$ of size $N_{\mathrm{BS}}\times N_{\mathrm{BS}}$ and $\mathbf{U}_{\mathrm{MS}}$ of size $N_{\mathrm{MS}}\times N_{\mathrm{MS}}$ as\footnote{In principle, the electromagnetic lens can be approximately modeled with \ac{DFT} matrices. An alternative approach is by integrating lens and antenna array as a single component \cite{ZengYy2}.} \cite{YongZen,Xinyu_Gau}
\begin{align}
\check{\mathbf{H}}_{\mathrm{DL}}[n]&=\mathbf{U}_{\mathrm{MS}}^{\mathrm{H}}\mathbf{H}_{\mathrm{DL}}[n]\mathbf{U}_{\mathrm{BS}}\label{BWTransceiver1a}\\
& = \sum_{k=0}^{K}\gamma_{n}(\tilde{h}_{k},\tau_{k})\mathbf{U}_{\mathrm{MS}}^{\mathrm{H}}\mathbf{a}_{\mathrm{MS}}(\phi_{k})\mathbf{a}^{\mathrm{H}}_{\mathrm{BS}}(\theta_{k})\mathbf{U}_{\mathrm{BS}},\label{BWTransceiver1az}
\end{align}
where $\mathbf{U}_{\mathrm{BS}}$ contains $N_{\mathrm{BS}}$ orthogonal beams covering the entire angular domain as
\begin{align}
\mathbf{U}_{\mathrm{BS}}&\triangleq\left[\mathbf{u}_{\mathrm{BS}}({-(N_{\mathrm{BS}}-1)/2}),\ldots,\mathbf{u}_{\mathrm{BS}}({(N_{\mathrm{BS}}-1)/2})\right],\label{BWTransceiver1abb}\\
\mathbf{u}_{\mathrm{BS}}(p)&\triangleq \begin{bmatrix}
e^{-j2\pi \frac{N_{\mathrm{BS}}-1}{2} \frac{p}{{N_{\mathrm{BS}}}}},\ldots,e^{j2\pi \frac{N_{\mathrm{BS}}-1}{2} \frac{p}{N_{\mathrm{BS}}}}
\end{bmatrix}^{\mathrm{T}},\label{BWTransceiver1acc}
\end{align}
and $\mathbf{U}_{\mathrm{MS}}$ is defined similarly by replacing the subscript BS with MS. It is readily verified that \cite{widebandbrady} 
\begin{align}
[\check{\mathbf{H}}[n]]_{\mathrm{DL},i,i'}&=\sum_{k=0}^{K}\gamma_{n}(\tilde{h}_{k},\tau_{k})\chi_{\mathrm{MS}}\big(\frac{d}{\lambda_{c}}\sin(\phi_{k})-\frac{i}{N_{\mathrm{MS}}}\big)\chi_{{\mathrm{BS}}}\big(\frac{d}{\lambda_{c}}\sin(\theta_{k})-\frac{i'}{N_{\mathrm{BS}}}\big),\label{BWTransceiver1b}
\end{align}
for $-(N_{\mathrm{MS}}-1)/2 \le i \le (N_{\mathrm{MS}}-1)/2$ and $-(N_{\mathrm{BS}}-1)/2 \le i' \le (N_{\mathrm{BS}}-1)/2$. We have introduced
\begin{align}
\chi_{\mathrm{BS}}(\phi) & = \frac{\sin(\pi N_{\mathrm{BS}}\theta)}{\sqrt{N_{\mathrm{BS}}}\sin(\pi\theta)},\label{chi_A}\\
\chi_{\mathrm{MS}}(\phi) & = \frac{\sin(\pi N_{\mathrm{MS}}\phi)}{\sqrt{N_{\mathrm{MS}}}\sin(\pi\phi)}.\label{chi_B}
\end{align}
Similarly, the uplink channel model in beamspace $\check{\mathbf{H}}_{\mathrm{UL}}[n]$ is obtained by interchanging $\mathbf{A}_{\mathrm{BS}}$ and $\mathbf{A}_{\mathrm{MS}}$ in \eqref{Channel1}, and $\mathbf{U}_{\mathrm{MS}}$ and $\mathbf{U}_{\mathrm{BS}}$ in \eqref{BWTransceiver1a}. 
\begin{figure}   
\psfrag{Block 0}{\tiny Block 0}
\psfrag{Block 1}{\tiny  Block 1}
\psfrag{Block 2}{\tiny Block 2}
\psfrag{AOA-TOA}{\tiny AOA-TOA}
\psfrag{AOA}{\tiny AOA}
\psfrag{uplink}{\tiny uplink}
\psfrag{downlink}{\tiny downlink}
\psfrag{TB}{\tiny $\color{red}T_{B}$}
\psfrag{tracking}{\tiny  tracking}
\psfrag{estimation}{\tiny estimation}
\psfrag{Data transmission}{\tiny Data transmission}
\psfrag{Ttrain}{\tiny $T_{\mathrm{train}}$}
\psfrag{Ttrack}{\tiny $T_{\mathrm{track}}$}
\centering
\includegraphics[width=0.9\columnwidth]{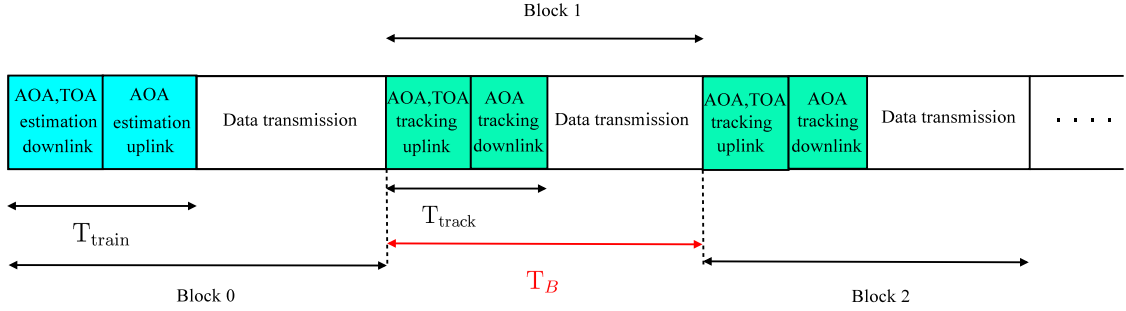}
  \caption{The illustration of a training-tracking system with reduced training time and significant reduction of tracking time in the uplink and downlink.}
  \label{Train_Track_Sch}
\end{figure}
\subsection{Received Signal in Beamspace}
The downlink received signal is obtained as
\begin{equation}\label{Receivedb1xc}
\check{\mathbf{y}}^{\mathrm{dl},(g)}[n]=(\mathbf{F}^{(g)}_{\mathrm{MS}})^{\mathrm{H}}\check{\mathbf{H}}_{\mathrm{DL}}[n]\mathbf{F}^{(g)}_{\mathrm{BS}}\mathbf{x}^{(g)}[n]+\check{\mathbf{n}}^{(g)}_{\mathrm{MS}}[n],
\end{equation}
where $\check{\mathbf{H}}_{\mathrm{DL}}[n]$ is the \ac{mm-wave} lens \ac{MIMO} channel in \eqref{BWTransceiver1a}, $\mathbf{F}^{(g)}_{\mathrm{MS}}=\mathbf{F}^{(g)}_{\mathrm{RF},\mathrm{MS}}\mathbf{F}_{\mathrm{BB},\mathrm{MS}}^{(g)}$ in which $\mathbf{F}_{\mathrm{BB},\mathrm{MS}}^{(g)}$ and $\mathbf{F}^{(g)}_{\mathrm{RF},\mathrm{MS}}$ are the digital baseband combiner and adaptive selecting network that is adaptively used as an analog combiner during the channel estimation. The adaptive selecting network is implemented using 1-bit phase shifters\footnote{The adaptive selecting network can be applied for traditional beam selection in data transmission by turning off some phase shifters and setting the other phase shifters to zero to realize ``unselec'' and ``select'', respectively. During beamspace channel estimation, it is adaptively used as an analog combiner to combine uplink or downlink signal.} for choosing the antenna index of the lens array \cite{Xinyu_Gau}. The noise vector is defined as $$\check{\mathbf{n}}^{(g)}_{\mathrm{MS}}[n]=(\mathbf{F}^{(g)}_{\mathrm{MS}})^{\mathrm{H}}\mathbf{U}_{\mathrm{MS}}^{\mathrm{H}}\mathbf{n}^{(g)}[n].$$ 
Finally, for $G$ sequentially transmitted beams we obtain the downlink received signal
\begin{equation}\label{BWTransceiver2x}
\check{\mathbf{y}}^{\mathrm{dl}}[n]=\mathbf{\Omega}[n]\check{\mathbf{h}}[n]+\check{\mathbf{n}}_{\mathrm{MS}}[n],
\end{equation}
where
\begin{align}
\mathbf{\Omega}[n]&=\begin{bmatrix}
\mathbf{\Omega}^{(1)}[n]\\\vdots\\\mathbf{\Omega}^{(G)}[n]
\end{bmatrix},\\
\mathbf{\Omega}^{(g)}[n]&=(\mathbf{Z}^{(g)}_{\mathrm{BS}}[n])^{\mathrm{T}}\otimes\left(\mathbf{F}_{\mathrm{MS}}^{(g)}\right)^{\mathrm{H}},\\
\mathbf{Z}^{(g)}_{\mathrm{BS}}[n]&=\mathbf{F}^{(g)}_{\mathrm{BS}}\mathbf{x}^{(g)}[n],\\
\check{\mathbf{h}}[n] &= \mathrm{vec}(\check{\mathbf{H}}_{\mathrm{DL}}[n]).
\end{align}
Similarly, the uplink received signal is obtained by replacing the uplink channel model in the beamspace $\check{\mathbf{H}}_{\mathrm{UL}}[n]=\mathbf{U}_{\mathrm{BS}}^{\mathrm{H}}\mathbf{H}_{\mathrm{UL}}[n]\mathbf{U}_{\mathrm{MS}}$ and interchanging $\mathbf{F}^{(g)}_{\mathrm{BS}}=\mathbf{F}^{(g)}_{\mathrm{RF},\mathrm{BS}}\mathbf{F}_{\mathrm{BB},\mathrm{BS}}^{(g)}$ and $\mathbf{F}^{(g)}_{\mathrm{MS}}=\mathbf{F}^{(g)}_{\mathrm{RF},\mathrm{MS}}\mathbf{F}_{\mathrm{BB},\mathrm{MS}}^{(g)}$ in \eqref{Receivedb1xc}. The terms $\mathbf{F}_{\mathrm{BB},\mathrm{BS}}^{(g)}$ and $\mathbf{F}^{(g)}_{\mathrm{RF},\mathrm{BS}}$ denote the digital baseband combiner and adaptive selecting network applied using 1-bit phase shifter in the \ac{BS}, respectively. Consequently, the received signal for the sequential transmission of $G$ signals in the uplink is obtained accordingly. Our goal is to provide a training-tracking system by uplink and downlink transmissions as shown in Fig. \ref{Train_Track_Sch}. In the training phase, i.e., Block $0$, the main aim is to reduce the required time $T_{\mathrm{train}}$ for channel estimation. This leads to saving more symbols for data transmission and consequently increasing the data rate \cite{GiusepJoint}. In the tracking phase, i.e., Block $m>0$, instead of re-estimating the channel with the required time $T_{\mathrm{train}}$, we propose an approach with significant reduction of training time, i.e. $T_{\mathrm{track}}\ll T_{\mathrm{train}}$. The duration of each block during the tracking phase is shown by\footnote{The value of $T_{B}$ can be set to $T_{B}\approx d_{0}/(v_{\mathrm{max}}N_{\mathrm{BS}})$ for the beamforming losses smaller than 3 dB at a given distance from the \ac{BS}, $d_{0}$, the maximum velocity $v_{\mathrm{max}}$, and antenna element spacing $d=\lambda_{c}/2$ for the worst case geometry \cite{Marzi}. Consequently, increasing $v_{\mathrm{max}}$ and $N_{\mathrm{BS}}$ leads to reducing $T_{B}$ while increasing $d_{0}$ results increasing $T_{B}$.} $T_{B}$. Moreover, the localization information from the tracking phase helps the data transmission in a position-aided communication system.
\section{Channel Estimation for Lens Antenna Array}\label{Pos_ Ori_Est}
A novel extended \ac{SD}-based method for frequency selective \ac{mm-wave} \ac{MIMO} channel is proposed for the estimation of channel parameters. In the training phase, it is assumed that the number of \ac{RF}-chains in the \ac{BS} and the \ac{MS} are set to one, i.e., $M_{\mathrm{BS}}=M_{\mathrm{MS}}=1$. The proposed method estimates the beamspace coefficients, \ac{AOA}, \ac{TOA}, and the number of channel paths $K$ in the downlink phase. In the uplink phase, the \ac{AOA} is estimated by the transmission of beams in the direction of the estimated \ac{AOA} obtained from the downlink. The method directly estimates the beamspace channel coefficients. The proposed method performs better than the conventional \ac{CS}-based approaches from the literature in terms of detection probability of the support of the sparse channel coefficient.
\subsection{Background on SD-Based Method}
The \ac{SD}-based channel estimation method is an approach for the estimation of the \ac{mm-wave} channel for the limited transmit power, i.e., low \ac{SNR} \cite{Xinyu_Gau}. In principle, the \ac{SD}-based channel estimation method is based on the following properties for the beamspace channel coefficients:
\begin{enumerate}
\item Beamspace channel coefficients for different paths are asymptotically orthogonal for sufficiently large number of antenna elements.
\item The ratio between the power of $V$ strongest elements of beamspace channel coefficient for a given path and the total power can be lower bounded. Moreover, the $V-1$ strongest elements are located uniformly around the strongest element of beamspace channel coefficient.
\end{enumerate}
Using the above properties, one can separately estimate the beamspace coefficients for different paths by starting from the strongest path and $V-1$ strongest elements of beamspace channel coefficient, removing their effects and repeating the process untill all the $\hat{K}+1$ paths are estimated. Using the second property, it is possible to retain a few strongest elements of the beamspace channel coefficients for each path and set the rest of elements to zero without considerable power loss. Considering the second property, the support of the sparse channel coefficient is obtained with higher accuracy using the \ac{SD}-based method compared to the traditional \ac{CS}-based methods. This is due to the fact that in the \ac{CS}-based methods only the position of strongest nonzero element of the sparse vector is obtained one by one for each path in an iterative way such that the magnitude and consequently the detection probability of strongest nonzero element for each path decreases \cite{Xinyu_Gau}. 
\subsection{Estimation of TOA/AOA and Beamspace Coefficients in the Downlink}\label{DL_ChEst}
To achieve desired recovery accuracy of the channel parameters for the training phase in the downlink, the adaptive selecting network in the \ac{MS}, $\mathbf{F}^{(g)}_{\mathrm{RF},\mathrm{MS}}$, is obtained using Bernoulli random matrix with the elements randomly selected from $\{-1,+1\}$ with equal probability for each sequentially transmitted symbol, indexed by $g$. The baseband precoder $\mathbf{F}^{(g)}_{\mathrm{BB},\mathrm{MS}}$ is populated with \ac{i.i.d.} complex entries on the unit circle. This makes the mutual coherence between different columns of the sensing matrix as small as possible. The \ac{BS} transmits equal power signal for all the elements of the lens antenna array. This is due to the fact that the \ac{AOD} in the downlink is not estimated in this stage and the best \ac{AOD} in the downlink to transmit highest signal power is not known a priori\footnote{This is equivalent to isotropical transmission \cite{Yang_Lu}.}. For equal power transmission in the \ac{BS} without prior location information, one can use the \ac{RF}-chain as a splitter\footnote{For the case of large number of antenna elements several splitters can be used for each subarray and the corresponding beams can be transmitted sequentially or simultaneously.}, i.e., $\mathbf{F}^{(g)}_{\mathrm{RF},\mathrm{BS}}=\mathbf{1}_{N_{\mathrm{BS}}}$, in the downlink. The received signal for the training phase in the downlink for the $g$-th sequentially transmitted symbol is simplified according to \eqref{Receivedb1xc} as
\begin{equation}\label{Training0}
\check{y}^{\mathrm{dl},(g)}[n]=x^{(g)}_{\mathrm{BB},\mathrm{BS}}[n](\mathbf{F}^{(g)}_{\mathrm{MS}})^{\mathrm{H}}\check{\mathbf{h}}_{\mathrm{MS}}[n]+\check{\mathbf{n}}^{(g)}_{\mathrm{MS}}[n],
\end{equation}
where\footnote{Without loss of generality, one can choose $x^{(g)}_{\mathrm{BB},\mathrm{BS}}[n]=1$ for all $g$. This simplifies the sensing matrix in \eqref{Training2} as $\bar{\mathbf{F}}_{\mathrm{MS}}$.} $x^{(g)}_{\mathrm{BB},\mathrm{BS}}[n]=F^{(g)}_{\mathrm{BB},\mathrm{BS}}x^{(g)}[n]$ is the precoded signal for each subcarrier $n$ and sequential index $g$. The term $\check{\mathbf{h}}_{\mathrm{MS}}[n]$ is defined as
\begin{equation}\label{Training6}
\check{\mathbf{h}}_{\mathrm{MS}}[n]=\sqrt{N_{\mathrm{BS}}}\sum_{k=0}^{K}\gamma_{n}(\tilde{h}_{k},\tau_{k})\boldsymbol{\chi}_{\mathrm{MS},k},
\end{equation}
in which $\boldsymbol{\chi}_{\mathrm{MS},k}$ is an $N_{\mathrm{MS}}\times 1$ vector with the entries $\chi_{\mathrm{MS}}\big(\frac{d}{\lambda_{c}}\sin(\phi_{k})-\frac{i}{N_{\mathrm{MS}}}\big)$ for $-(N_{\mathrm{MS}}-1)/2\leq i\leq (N_{\mathrm{MS}}-1)/2$. In computing \eqref{Training0}, we used the fact that 
\begin{equation}\label{Training5}
\boldsymbol{\chi}^{\mathrm{T}}_{\mathrm{BS},k}\mathbf{1}_{N_{\mathrm{BS}}}=\sum_{i'=-(N_{\mathrm{BS}}-1)/2}^{(N_{\mathrm{BS}}-1)/2}\chi_{{\mathrm{BS}}}\big(\frac{d}{\lambda_{c}}\sin(\theta_{k})-\frac{i'}{N_{\mathrm{BS}}}\big)\approx\sqrt{N_{\mathrm{BS}}},\:\forall\theta_{k}\in [0, 2\pi)
\end{equation}
where $\boldsymbol{\chi}_{\mathrm{BS},k}$ is an $N_{\mathrm{BS}}\times 1$ vector with the entries $\chi_{\mathrm{BS}}\big(\frac{d}{\lambda_{c}}\sin(\theta_{k})-\frac{i'}{N_{\mathrm{BS}}}\big)$ for $-(N_{\mathrm{BS}}-1)/2\leq i'\leq (N_{\mathrm{BS}}-1)/2$. Consequently, for $G$ sequentially transmitted beams we obtain the $G\times 1$ received signal 
\begin{equation}\label{Training1}
\check{\mathbf{y}}^{\mathrm{dl}}[n]=\mathbf{\Omega}_{\mathrm{MS}}^{\mathrm{H}}[n]\check{\mathbf{h}}_{\mathrm{MS}}[n]+\check{\mathbf{n}}_{\mathrm{MS}}[n],
\end{equation}
where
\begin{equation}\label{Training2}
\mathbf{\Omega}_{\mathrm{MS}}[n]=\bar{\mathbf{F}}_{\mathrm{MS}}\mathrm{diag}\left\{\left(x^{(1)}_{\mathrm{BB},\mathrm{BS}}[n]\right)^{*},\ldots,\left(x^{(G)}_{\mathrm{BB},\mathrm{BS}}[n]\right)^{*}\right\},
\end{equation}\label{Training3}
where $\bar{\mathbf{F}}_{\mathrm{MS}}=\begin{bmatrix}\mathbf{F}^{(1)}_{\mathrm{MS}},\ldots,\mathbf{F}^{(G)}_{\mathrm{MS}}\end{bmatrix}$ denotes the $N_{\mathrm{MS}}\times G$ sensing matrix with the $i$-th column vector $\mathbf{F}^{(i)}_{\mathrm{MS}}$. Since there exist only a few non-zero elements in \eqref{Training6} for each path, $\check{\mathbf{h}}_{\mathrm{MS}}[n]$ is a sparse vector. Consequently, the problem \eqref{Training1} can be considered as a sparse signal recovery with the sensing matrix defined in \eqref{Training2}. Moreover, since $\bar{\mathbf{F}}_{\mathrm{MS}}$ is designed to exhibit low mutual coherence among its columns, by choosing random phase shifters for both \ac{RF} and baseband parts, it is guaranteed that the complete information is available even for the values of $G$ much less than $N_{\mathrm{MS}}$. In particular to solve \eqref{Training1}, the number of sequentially transmitted signals\footnote{For compactness, Hardy's notation is used \cite{BajwaWzzd}.} should satisfy $G\succeq (\hat{K}+1)\log (N_{\mathrm{MS}}/(\hat{K}+1))$ \cite{BajwaWzzd}.

The channel estimation in the downlink is explained in Algorithm \ref{algor0_det}. The received signal in the downlink $\check{\mathbf{y}}^{\mathrm{dl}}[n]$ in \eqref{Training1}, sensing matrix $\mathbf{\Omega}_{\mathrm{MS}}[n]$ in \eqref{Training2}, the required number of strongest elements $V$, and the threshold $\delta$ are the input parameters. The output parameters are the estimates of $K$, $\phi_{k}$, and $\check{\mathbf{h}}_{\mathrm{MS}}[n]$ for $n=0,\ldots,N-1$. First, the values of the residual vector $\mathbf{r}^{\mathrm{dl}}_{t}[n]$ and beamspace channel coefficient $\check{\mathbf{h}}_{\mathrm{MS},t}[n]$ are initialized as, $\mathbf{r}^{\mathrm{dl}}_{-1}[n]=\mathbf{0}_{N_{\mathrm{MS}}}$, $\mathbf{r}^{\mathrm{dl}}_{0}[n]=\check{\mathbf{y}}^{\mathrm{dl}}[n]$, and $\check{\mathbf{h}}_{\mathrm{MS},t}[n]=\mathbf{0}_{N_{\mathrm{MS}}}$, respectively. The corresponding index of the $t$-th \ac{AOA} is estimated by finding the maximum value of the inner product of the $m_{\mathrm{MS}}$-th row of sensing matrix $\mathbf{\Omega}_{\mathrm{MS}}[n]$ defined as $\boldsymbol{\omega}_{\mathrm{MS},m_{\mathrm{MS}}}$ and the residual vector $\mathbf{r}^{\mathrm{dl}}_{t}[n]$ for $m_{\mathrm{MS}}=1,\ldots,N_{\mathrm{MS}}$ in \eqref{Training15}. Consequently, the \ac{AOA} is obtained based on \eqref{Training16}. The support region for the corresponding index $\tilde{m}_{\mathrm{MS},t}$ is computed based on \ac{SD}-based approach for the even number of strongest elements\footnote{For the odd number of strongest elements $V$ and the $t$-th path, the support would be $\mathrm{supp}\left(\boldsymbol{\chi}_{\mathrm{MS},t}\right)=\mathrm{mod}_{N_{\mathrm{MS}}}\left\{\tilde{m}_{\mathrm{MS},t}-\frac{V-1}{2},\ldots,\tilde{m}_{\mathrm{MS},t}+\frac{V-1}{2}\right\}$. In principle, the value of $V$ is not known a priori and can be set to a fixed number for a given energy detection ratio.} $V$ in \eqref{Training17} where the operation $\mathrm{mod}_{N_{\mathrm{MS}}}\left\{.\right\}$ ensures the support not to exceed $N_{\mathrm{MS}}$. Next, the beamspace channel coefficient for the $t$-th path $\check{\mathbf{h}}_{\mathrm{MS},t}[n]$ is computed based on the \ac{LS} solution for \eqref{Training1} obtained as \eqref{Training18}. Consequently, the residual vector $\mathbf{r}^{\mathrm{dl}}_{t-1}[n]$ is updated by subtracting the effect of corresponding path from the previous residual vector $\mathbf{r}^{\mathrm{dl}}_{t-1}[n]$ in \eqref{Training19} and the iteration index $t$ is increased by 1. Finally, the aforementioned steps are repeated until the stopping criteria defined as $\sum_{n=0}^{N-1}\Vert\mathbf{r}^{\mathrm{dl}}_{t-1}[n]-\mathbf{r}^{\mathrm{dl}}_{t-2}[n]\Vert^{2}_{2}>\delta$ is fulfilled for a given threshold $\delta$. The threshold $\delta$ is obtained using \ac{CFAR} test with a similar procedure as in \cite{Marzi}:
\begin{equation}\label{Threshold_express}
\delta=N_{0}\gamma^{-1}\left(N,\Gamma\left(N\right)\left(1-\mathbb{P}_{\mathrm{fa}}\right)^{1/N_{\mathrm{MS}}}\right),
\end{equation}
where $\gamma^{-1}\left(N,x\right)$ denotes the inverse of the incomplete gamma distribution, $\Gamma(N)$ is the gamma function, and $\mathbb{P}_{\mathrm{fa}}$ is the false alarm probability. Using the estimated values of the beamspace coefficients $\check{\mathbf{h}}_{\mathrm{MS},t}[n]$ for $n=0,\ldots,N-1$, $\gamma_{n}(\tilde{h}_{k},\tau_{k})$ and consequently $\tilde{h}_{k}$ and $\tau_{k}$ are estimated for each path $k=0,\ldots,\hat{K}$ based on a similar \ac{LS} solution as in \cite{Arash2x}.
\begin{algorithm}
    \SetKwInOut{Input}{Input}
    \SetKwInOut{Output}{Output}
    \Input{Received signals $\check{\mathbf{y}}^{\mathrm{dl}}[n]$, sensing matrix $\mathbf{\Omega}_{\mathrm{MS}}[n]$, required number of strongest elements $V$, and the threshold $\delta$.}
    \Output{estimates of $K$, $\phi_{k}$, and $\hat{\check{\mathbf{h}}}_{\mathrm{MS},k}[n]\big\vert_{\mathrm{supp}\left(\boldsymbol{\chi}_{\mathrm{MS},k}\right)}$ for $n=0,\ldots,N-1$.}
    \textbf{Initialization:}
    For $n=0,\ldots,N-1$ and $t=0$, the residual vectors and beamspace coefficients are set to $\mathbf{r}^{\mathrm{dl}}_{-1}[n]=\mathbf{0}_{N_{\mathrm{MS}}}$, $\mathbf{r}^{\mathrm{dl}}_{0}[n]=\check{\mathbf{y}}^{\mathrm{dl}}[n]$, and $\hat{\check{\mathbf{h}}}_{\mathrm{MS},0}[n]=\mathbf{0}_{N_{\mathrm{MS}}}$\;
    \While{$\sum_{n=0}^{N-1}\Vert\mathbf{r}^{\mathrm{dl}}_{t-1}[n]-\mathbf{r}^{\mathrm{dl}}_{t-2}[n]\Vert^{2}_{2}>\delta$}{
     \BlankLine
    \begin{itemize}
    \item Find \ac{AOA} in the downlink 
    \end{itemize}
    \begin{align}
    \tilde{m}_{\mathrm{MS},t}& =\underset{m_{\mathrm{MS}}=1,\ldots,N_{\mathrm{MS}}}
    {\mathrm{argmax}} \:\sum_{n=0}^{N-1}\frac{\vert 
   \boldsymbol{\omega}_{\mathrm{MS},m_{\mathrm{MS}}}^{\mathrm{H}}[n]\mathbf{r}^{\mathrm{dl}}_{t-1}[n]\vert}{\Vert        \bar{\mathbf{f}}_{\mathrm{MS},x,m_{\mathrm{MS}}}[n]\Vert_{2}},\label{Training15}\\
     \hat{\phi}_{t}& =\arcsin\left((\lambda_{c}/d)(\tilde{m}_{\mathrm{MS},t}-{(N_{\mathrm{MS}}-1)/2}-1)/N_{\mathrm{MS}}\right).\label{Training16}
     \end{align}
      \begin{itemize}
    \item Compute the sparsity support with the even number of elements $V$ as
    \end{itemize}
    \begin{equation}\label{Training17}
     \mathrm{supp}\left(\boldsymbol{\chi}_{\mathrm{MS},t}\right)= \mathrm{mod}_{N_{\mathrm{MS}}}\left\{\tilde{m}   _{\mathrm{MS},t}-\frac{V}{2},\ldots,\tilde{m}_{\mathrm{MS},t}+\frac{V-2}{2}\right\}.
      \end{equation}
       \begin{itemize}
       \item Compute $\hat{\check{\mathbf{h}}}_{\mathrm{MS},t}[n]$ for $n=0,\ldots,N-1$ as
       \end{itemize}
     \begin{equation}\label{Training18}
     \hat{\check{\mathbf{h}}}_{\mathrm{MS},t}[n]\big\vert_{\mathrm{supp}\left(\boldsymbol{\chi}_{\mathrm{MS},t}\right)}=   \left[\mathbf{\Omega}_{\mathrm{MS}}\right]_{(:,\mathrm{supp}\left(\boldsymbol{\chi}_{\mathrm{MS},t}\right))}  ^{\dagger}[n]\mathbf{r}^{\mathrm{dl}}_{t-1}[n].
     \end{equation}
     \begin{itemize}
     \item Update the residual vector $\mathbf{r}^{\mathrm{dl}}_{t}[n]$  for $n=0,\ldots,N-1$ as
     \end{itemize}
    \begin{equation}\label{Training19}
    \mathbf{r}^{\mathrm{dl}}_{t}[n]=\mathbf{r}^{\mathrm{dl}}_{t-1}[n]-\mathbf{\Omega}_{\mathrm{MS}}[n]   \hat{\check{\mathbf{h}}}_{\mathrm{MS},t}[n]\big\vert_{\mathrm{supp}\left(\boldsymbol{\chi}_{\mathrm{MS},t}\right)}.
     \end{equation}
     \begin{itemize}
     \item Set $t=t+1$.
     \end{itemize}
    }
    \caption{Extended \ac{SD}-Based Channel Estimation in the Downlink\label{algor0_det}}
\end{algorithm}
\subsection{Estimation of AOA in the Uplink}
In the uplink, the \ac{AOA} is estimated\footnote{To reduce the complexity, it is assumed that there is no clock bias in the \ac{BS} or \ac{MS}, consequently the \ac{TOA} is only estimated in the downlink with no need for bias compensation using two-way \ac{TOA} estimation protocols.} by the sequential transmission with one \ac{RF}-chain in the \ac{MS},i.e., $M_{\mathrm{MS}}=1$. In the uplink the \ac{BS} has no knowledge of the \ac{AOA} and applies the analog combiner $\mathbf{F}^{(g)}_{\mathrm{RF},\mathrm{BS}}$ using the Bernoulli random matrix with the elements randomly selected from $\{-1,+1\}$ with equal probability. Similar to the downlink, $\mathbf{F}^{(g)}_{\mathrm{BB},\mathrm{BS}}$ is set to \ac{i.i.d.} complex entries populated on the unit circle. This makes the mutual coherence between different columns of the sensing matrix as small as possible similar to the downlink. The received signal for the training phase in the uplink is simplified as
\begin{equation}\label{Training20}
\check{\mathbf{y}}^{\mathrm{ul}}[n]=\mathbf{\Omega}_{\mathrm{BS}}^{\mathrm{H}}[n]\check{\mathbf{h}}_{\mathrm{BS}}[n]+\check{\mathbf{n}}_{\mathrm{BS}}[n],
\end{equation}
where
\begin{equation}\label{Training21}
\mathbf{\Omega}_{\mathrm{BS}}[n]=\bar{\mathbf{F}}_{\mathrm{BS}}\mathrm{diag}\left\{\left(x^{(1)}_{\mathrm{BB},\mathrm{MS}}[n]\right)^{*},\ldots,\left(x^{(G)}_{\mathrm{BB},\mathrm{MS}}[n]\right)^{*}\right\},
\end{equation}\label{Training22}
in which $\bar{\mathbf{F}}_{\mathrm{BS}}=\begin{bmatrix}\mathbf{F}^{(1)}_{\mathrm{BS}},\ldots,\mathbf{F}^{(G)}_{\mathrm{BS}}\end{bmatrix}$ denotes the $N_{\mathrm{BS}}\times G$ sensing matrix with the $i$-th column vector $\mathbf{F}^{(i)}_{\mathrm{BS}}$, and $x^{(g)}_{\mathrm{BB},\mathrm{MS}}[n]=F^{(g)}_{\mathrm{BB},\mathrm{MS}}x^{(g)}[n]$ is the precoded signal in the \ac{MS}. The term $\check{\mathbf{h}}_{\mathrm{BS}}[n]$ in \eqref{Training20} denotes
\begin{equation}\label{Training23}
\check{\mathbf{h}}_{\mathrm{BS}}[n]=\sqrt{N_{\mathrm{MS}}}\sum_{k=0}^{K}\gamma_{n}(\tilde{h}_{k},\tau_{k})\boldsymbol{\chi}_{\mathrm{BS},k},
\end{equation}
where we used a similar principle as in \eqref{Training5} using an all-one precoder in the \ac{MS}. The problem \eqref{Training20} is considered as a sparse signal recovery with the sensing matrix defined in \eqref{Training21}, since there exist only a few non-zero elements in \eqref{Training23} for each path.
Using the received signal in the uplink $\check{\mathbf{y}}^{\mathrm{ul}}[n]$ in \eqref{Training20}, the \ac{AOA}s in the uplink are estimated with the same principle as in Section \ref{DL_ChEst}, for $\hat{K}+1$ paths accounting for all subcarriers\footnote{In principle, the angular sparsity pattern does not change for different subcarriers provided that the conditions in\cite{widebandbrady,Arash2x} hold. Consequently, it is sufficient to consider a few subcarriers in the uplink in practice. Nevertheless, if this is not the case due to beam squint, the Algorithm 1 needs to be repeated for each subcarrier. A procedure for an antenna array at the \ac{BS} and a single antenna at the \ac{MS} considering the effect of beam squint was proposed in \cite{XinyuBb}.}. Moreover, the angular refinement can be achieved using a similar procedure as in \cite{Arash2x}. Finally, the proposed method can be easily extended for the multiuser channel training by repeating the same precedures with randomly selected signals in the uplink and downlink for each user.

The complexity analysis of different steps of the training phase is proposed in the Appendix \ref{app_A}.
\subsection{Required Time for Training}
The required time for channel training using the proposed method is significantly reduced compared to traditional methods. The traditional methods including exhaustive sequential transmission in $N_{\mathrm{BS}}$ directions and exhaustive sequential reception in $N_{\mathrm{MS}}$ directions, and hierarchical codebook designs with progressively broader beams with $N_{\mathrm{RF}}$ \ac{RF} chains require the training times of $T_{\mathrm{train}}=T_{s}N_{\mathrm{BS}}N_{\mathrm{MS}}$ and $T_{\mathrm{train}}=T_{s}M_{G}(\hat{K}+1)^{2}\lceil\frac{M_{G}(\hat{K}+1)}{N_{\mathrm{RF}}}\rceil\log^{\frac{N_{G}}{\hat{K}+1}}_{M_{G}}$ where $N_{G}$ is the term denoting the grid resolution $2\pi/N_{G}$ and $M_{G}$ denotes the number of precoders, respectively \cite{AlkhateebA}.
On the contrary, the proposed method requires the training time of $GT_{s}$ in the downlink and $GT_{s}$ in the uplink and the total training time of $T_{\mathrm{train}}=2GT_{s}$ where $G\succeq (\hat{K}+1)\log (N_{\mathrm{BS}/\mathrm{MS}}/(\hat{K}+1))$. Moreover, the detection probability of the support of the sparse channel coefficient (i.e., the estimation accuracy) of the proposed method is better than the corresponding adopted \ac{CS}-based solutions for lens antenna arrays especially for low \ac{SNR} as will be explained in the simulations.
Following the above discussion, one can save several symbols from channel training for different purposes during the \ac{mm-wave} channel coherence time\footnote{The time that \ac{mm-wave} channel does not change (i.e., coherence time) is usually small due to large $f_{c}$. However, it includes several symbols thanks to the large bandwidth.}, e.g., communication, localization, and so on. After changing channel parameters due to different reasons, e.g., user movement, the estimated values from the training phase are not valid anymore. In this case, instead of re-estimating channel parameters with small changes that is complex and costly, we propose a tracking phase in the next section.  
\section{Tracking of Channel Parameters and Position and Orientation}\label{Pos_ Ori_Estb}
In this section we propose a tracking approach based on the \ac{EKF} method for channel parameters and position and orientation based on the information provided by the \ac{LOS} link\footnote{In the tracking phase, one can use the estimated channel parameters from the NLOS links obtained from the training phase for location estimation of stationary first-order reflectors, i.e., mapping the environment \cite{Arash2x}.}. It is assumed that the \ac{BS} does not move and tracks the location of the \ac{MS}, and the \ac{MS} tracks its rotation angle using the location information provided by the \ac{BS}.
\subsection{Measurement and State Equations}
A \ac{CWNA} model is employed for the state evolution for tracking \ac{AOA}/\ac{AOD}, and \ac{TOA} \cite{Bar_ShalomY}. The state vector for the \ac{LOS} path can be written as\footnote{Note that for tracking user location and orientation channel coefficient $\tilde{h}_{0}$ acts as a nuisance parameter and can be omitted from the state vector to reduce the computational complexity. This is due to the fact that \ac{AOA}/\ac{AOD} and \ac{TOA} obtained from the \ac{LOS} are sufficient for position and orientation estimation. Equivalently, the state vector can be written for position and rotation angle and converted to the channel parameters after tracking.}
\begin{equation}\label{track1}
\boldsymbol{\psi}_{0}^{[m]}=\begin{bmatrix}(\boldsymbol{\eta}_{0}^{[m]})^{\mathrm{T}}& (\dot{\boldsymbol{\eta}}_{0}^{[m]})^{\mathrm{T}}\end{bmatrix}^{\mathrm{T}},
\end{equation}
where $\boldsymbol{\eta}_{0}^{[m]}=[\tau_{0}^{[m]},\theta_{0}^{[m]},\phi_{0}^{[m]}]^{\mathrm{T}}$ and $\dot{\boldsymbol{\eta}}_{0}^{[m]}=[\dot\tau_{0}^{[m]},\dot\theta_{0}^{[m]},\dot\phi_{0}^{[m]}]^{\mathrm{T}}$. The terms $\theta_{0}^{[m]}$ and $\phi_{0}^{[m]}$ denote the \ac{AOD} and \ac{AOA} for the \ac{LOS} path at the time instant $m$, respectively. Similarly, $\tau_{0}^{[m]}$ denotes the \ac{TOA} for the \ac{LOS} path. Finally, the parameters $\dot\tau_{0}^{[m]}$, $\dot\theta_{0}^{[m]}$, and $\dot\phi_{0}^{[m]}$ denote the rate-of-change of the \ac{TOA}, \ac{AOD}, and \ac{AOA} for the block duration $T_{B}$, respectively. Assuming \ac{CWNA} model, the state evolution model can be written as
\begin{equation}\label{track2}
\boldsymbol{\psi}_{0}^{[m]}=\mathbf{\Phi}\boldsymbol{\psi}_{0}^{[m-1]}+\mathbf{u}_{0}^{[m]},
\end{equation}
where $\mathbf{u}_{0}^{[m]}$ denotes the state noise with $\mathbb{E}\left[\mathbf{u}_{0}^{[m]}(\mathbf{u}_{0}^{[m]})^{\mathrm{T}}\right]=\mathbf{Q}_{0}^{[m]}$. In general, the bi-azimuth generalized \ac{VMF} distribution for joint \ac{AOA}/\ac{AOD} or its approximation by a 2-D truncated Gaussian pdf for slightly distributed path components can be applied for directional data \cite{CzinkN,YinX}. In this case, we applied the approximation with a 2-D truncated Gaussian pdf with
$\sigma_{\phi_{0}}$, $\sigma_{\theta_{0}}$, and $\rho_{\theta\phi,0}$ denoting the direction spreads of the \ac{AOA}, \ac{AOD}, and cross correlation for the \ac{LOS} path, respectively. Moreover, the amount of noise would depend on $T_{B}$, the time between pilot transmissions. The state transition matrix $\mathbf{\Phi}\in\mathbb{R}^{6\times 6}$ is defined as
\begin{IEEEeqnarray}{rCl}
\mathbf{\Phi} & = & \begin{bmatrix}
\mathbf{I}_{3} & T_{B}\mathbf{I}_{3}\\
\mathbf{0}_{3} & \mathbf{I}_{3}
\end{bmatrix}.\label{track3}
\end{IEEEeqnarray}
For $m=1$, the entries of $\boldsymbol{\psi}_{0}^{[m-1]}$ in \eqref{track2} are initialized as: $\tau_{0}^{[0]}=\hat{\tau}_{0}$, $\phi_{0}^{[0]}=\hat{\phi}_{0}$, and $\theta_{0}^{[0]}=\hat{\theta}_{0}$ where $\hat{\tau}_{0}$, $\hat{\phi}_{0}$, and $\hat{\theta}_{0}$ are obtained from the training phase. The rate-of-change terms are initialized\footnote{The acceleration terms $\ddot{\theta}_{0}^{[m]}$, $\ddot{\theta}_{0}^{[m]}$, and $\ddot{\phi}_{0}^{[m]}$ can be incorporated in \eqref{track1} and defined as $\ddot{\tau}_{0}^{[m]}=(\dot\tau_{0}^{[m]}-\dot\tau_{0}^{[m-1]})/T_{B}$, $\ddot{\theta}_{0}^{[m]}=(\dot\theta_{0}^{[m]}-\dot\theta_{0}^{[m-1]})/T_{B}$, and $\ddot{\phi}_{0}^{[m]}=(\dot\phi_{0}^{[m]}-\dot\phi_{0}^{[m-1]})/T_{B}$, respectively, and initialized similarly.} by two consecutive estimates of $\boldsymbol{\eta}_{0}[m]=[\tau^{[m]}_{0},\theta^{[m]}_{0},\phi^{[m]}_{0}]^{\mathrm{T}}$ as: $\dot\tau_{0}^{[1]}=(\tau_{0}^{[1]}-\tau_{0}^{[0]})/T_{B}$, $\dot\theta_{0}^{[1]}=(\theta_{0}^{[1]}-\theta_{0}^{[0]})/T_{B}$, and $\dot\phi_{0}^{[1]}=(\phi_{0}^{[1]}-\phi_{0}^{[0]})/T_{B}$.
For tracking of the channel parameters, the \ac{EKF} is applied with the state comprising the \ac{LOS} delay, \ac{DL-AOA}, \ac{UL-AOA}, and their corresponding rates of changes, with the linear process model and nonlinear measurement equations in the downlink and the uplink. The measurement equation in the uplink\footnote{If the rates of change for the \ac{AOA} and \ac{AOD} during the tracking phase are equal, then it is sufficient to use only the uplink transmission as the measurement equation. This is due to the fact that the rotation angle can be obtained as a direct consequence of \eqref{geom_track2}.} is obtained for all the subcarriers (with the designed beamformers in the \ac{BS} and the \ac{MS} directed towards the \ac{LOS} link as will be explained in the next section) as
\begin{equation}\label{track6}
\check{\mathbf{y}}^{\mathrm{ul}, [m]}_{r}=\boldsymbol{\vartheta}^{\mathrm{ul}}_{0}(\boldsymbol{\eta}_{0}^{[m]})+\sum_{l=1}^{\hat{K}} \boldsymbol{\vartheta}^{\mathrm{ul}}_{0}(\boldsymbol{\eta}_{l}^{[m]})+\check{\mathbf{n}}_{\mathrm{BS}}^{[m]}.
\end{equation}
In \eqref{track6}, the second term denotes the superposition of all the other \ac{NLOS} paths acting as an added term to the measurement noise $\check{\mathbf{n}}_{\mathrm{BS}}^{[m]}=\mathbf{F}^{\mathrm{H}}_{\mathrm{BS},0}\check{\mathbf{n}}^{[m]}$ where $\mathbf{F}_{\mathrm{BS},0}$ is obtained based on the robust/heuristic design explained in the next section. The \ac{TOA}-\ac{AOA} are the parameters to be tracked for the block index $m$ in the \ac{BS}, and $\boldsymbol{\vartheta}^{\mathrm{ul}}_{0}(\boldsymbol{\eta}_{l}^{[m]})$ denotes
\begin{equation}\label{track7}
\boldsymbol{\vartheta}^{\mathrm{ul}}_{0}(\boldsymbol{\eta}_{k}^{[m]})=\tilde{h}_{k}^{[m]}\left(\mathbf{X}^{\mathrm{T}}_{0}(\mathbf{F}^{[m-1]}_{\mathrm{MS},0})^{\mathrm{T}}\boldsymbol{\chi}_{\mathrm{MS},k}^{[m]}\odot\mathbf{a}(\tau_{k}^{[m]})\right)\otimes\mathbf{F}^{\mathrm{H}}_{\mathrm{BS},0}\boldsymbol{\chi}_{\mathrm{BS},k}^{[m]},
\end{equation}
where $\mathbf{F}^{[m-1]}_{\mathrm{MS},0}$ denotes the uplink beamformer designed based on $\hat{\phi}_{0}^{[m-1]}$ and its uncertainty obtained from the covariance of the corresponding term from the \ac{EKF}, i.e., the first diagonal element of the covariance estimate of $\boldsymbol{\psi}_{\mathrm{dl},0}^{[m]}=[\phi_{0}^{[m]}, \dot{\phi}_{0}^{[m]}]^{\mathrm{T}}$ denoted by $[\mathbf{P}_{\boldsymbol{\psi}_{\mathrm{dl},0}}^{[m-1\vert m-1]}]_{1,1}$ as will be described in Section \ref{Rob_Heur}. The terms $\boldsymbol{\chi}_{\mathrm{BS},k}^{[m]}$ and $\boldsymbol{\chi}_{\mathrm{MS},k}^{[m]}$ denote the vectors $\boldsymbol{\chi}_{\mathrm{BS},k}$ and $\boldsymbol{\chi}_{\mathrm{MS},k}$ evaluated at $\theta_{k}^{[m]}$ and $\phi_{k}^{[m]}$, respectively. The delay vector is defined as $\mathbf{a}(\tau_{k})=[1,\ldots,e^{-j2\pi (N-1)\tau_{k}/(NT_{s})}]^{\mathrm{T}}$, and $\mathbf{X}_{0}=[\mathbf{x}^{(0)}[0],\ldots,\mathbf{x}^{(0)}[N-1]]^{\mathrm{T}}$ where $\mathbf{x}^{(0)}[n]$ is the $M_{\mathrm{MS}}\times 1$ vector of simultaneously transmitted symbols for the $n$-th subcarrier for the \ac{LOS} link.
Similarly, the measurement equation in the downlink is obtained for all the subcarriers for the designed beamformers towards the \ac{LOS} as
\begin{equation}\label{track8b}
\check{\mathbf{y}}^{\mathrm{dl}, [m]}_{r}=\boldsymbol{\vartheta}^{\mathrm{dl}}_{0}(\boldsymbol{\eta}_{0}^{[m]})+\sum_{l=1}^{\hat{K}} \boldsymbol{\vartheta}^{\mathrm{dl}}_{0}(\boldsymbol{\eta}_{l}^{[m]})+\check{\mathbf{n}}_{\mathrm{MS}}^{[m]},
\end{equation}
where $\boldsymbol{\vartheta}^{\mathrm{dl}}_{0}(\boldsymbol{\eta}_{l}^{[m]})$ is obtained similar to \eqref{track7}, by replacing the subscripts \ac{MS} and \ac{BS} with \ac{BS} and \ac{MS}, respectively. The term $\phi_{0}^{[m]}$ is the parameter to be tracked in the \ac{MS}, and the downlink beamformer $\mathbf{F}^{[m-1]}_{\mathrm{BS},0}$ is designed based on $\hat{\theta}_{0}^{[m-1]}$ and its uncertainty obtained from the covariance of the corresponding term from the \ac{EKF}, i.e., the second diagonal element of the covariance estimate of $\boldsymbol{\psi}_{\mathrm{ul},0}^{[m]}=[\tau_{0}^{[m]}, \theta_{0}^{[m]}, \dot{\tau}_{0}^{[m]}, \dot{\theta}^{[m]}]^{\mathrm{T}}$ denoted by $[\mathbf{P}_{\boldsymbol{\psi}_{\mathrm{ul},0}}^{[m-1\vert m-1]}]_{2,2}$ detailed in Section \ref{Rob_Heur}. In summary, the tracking phase starts in the uplink by transmission of the beams towards the \ac{LOS} using the \ac{AOA} information obtained from the training phase in the downlink, then the \ac{TOA}/\ac{AOA} are tracked at the \ac{BS}. Using the \ac{TOA}/\ac{AOA} in the uplink, the \ac{BS} computes the location of the \ac{MS}. Finally, using the location information and tracked \ac{AOA} in the downlink, the \ac{MS} tracks its rotation angle. The procedure is repeated within the observation time based on the information obtained from the previous tracking block.
\subsection{Conversion to Position and Orientation}
Using the channel parameters after tracking in $\{\hat{\tau}_{0}^{[m]},\hat{\theta}_{0}^{[m]},\hat{\phi}_{0}^{[m]}\}$, the tracked values of position and rotation angle $\left\{\hat{\mathbf{p}}^{[m]},\hat{\alpha}^{[m]}\right\}$ can be obtained using the following equations for the \ac{LOS} path
\begin{IEEEeqnarray}{rCl}
\hat{\mathbf{p}}^{[m]} & = & \mathbf{q}+c\hat{\tau}_{0}^{[m]}\left[\cos(\hat{\theta}_{0}^{[m]}), \sin(\hat{\theta}_{0}^{[m]})\right]^{\mathrm{T}},\label{geom_track1}\\
\hat{\alpha}^{[m]} & = & \pi+\hat{\theta}_{0}^{[m]}-\hat{\phi}_{0}^{[m]}.\label{geom_track2}
\end{IEEEeqnarray}
The uncertainty of position $\mathbf{P}^{[m\vert m]}_{\mathbf{p}}$ and rotation angle $P^{[m\vert m]}_{\alpha}$ are computed as
\begin{IEEEeqnarray}{rCl}
\mathbf{P}^{[m\vert m]}_{\mathbf{p}} &=& \left(\mathbf{T}(\hat{\tau}_{0}^{[m]},\hat{\theta}_{0}^{[m]})\left[\mathbf{P}_{\boldsymbol{\psi}_{\mathrm{ul},0}}^{[m\vert m]}\right]_{1:2,1:2}^{-1}\mathbf{T}^{\mathrm{T}}(\hat{\tau}_{0}^{[m]},\hat{\theta}_{0}^{[m]})\right)^{-1},\label{geom_track4}\\
P^{[m\vert m]}_{\alpha}&=&\left[\mathbf{P}_{\boldsymbol{\psi}_{\mathrm{ul},0}}^{[m\vert m]}\right]_{2,2}+\left[\mathbf{P}_{\boldsymbol{\psi}_{\mathrm{dl},0}}^{[m\vert m]}\right]_{1,1},\label{geom_track5}
\end{IEEEeqnarray}
where $\mathbf{T}(\hat{\tau}_{0}^{[m]},\hat{\theta}_{0}^{[m]})$ denotes the conversion matrix $\mathbf{T}(\tau_{0},\theta_{0})=[\frac{\partial\tau_{0}}{\partial\mathbf{p}}, \frac{\partial\theta_{0}}{\partial\mathbf{p}}]$ evaluated at $(\hat{\tau}_{0}^{[m]},\hat{\theta}_{0}^{[m]})$ with 
\begin{IEEEeqnarray}{rCl}
\frac{\partial\tau_{0}}{\partial\mathbf{p}} & = & \frac{1}{c}\left[\cos(\theta_{0}), \sin(\theta_{0})\right]^{\mathrm{T}}, \label{geom_track6}\\
\frac{\partial\theta_{0}}{\partial\mathbf{p}} & = & \frac{1}{c\tau_{0}}\left[-\sin(\theta_{0}), \cos(\theta_{0})\right]^{\mathrm{T}}.
\end{IEEEeqnarray}
The operation $\left[.\right]^{-1}_{1:2,1:2}$ in \eqref{geom_track4} denotes the $2\times 2$ upper left submatrix of the inverse of the argument.
%
%
%
In principle, the information from the \ac{LOS} link (i.e., equations \eqref{geom_track1}-\eqref{geom_track2}) is sufficient to track user position and rotation angle $\left\{\hat{\mathbf{p}}^{[m]},\hat{\alpha}^{[m]}\right\}$ with an acceptable accuracy. Moreover, gaining the \ac{NLOS} links information to improve position and rotation angle accuracies in the tracking phase requires \emph{simultaneous} transmission of the beams in the \ac{LOS} and \ac{NLOS} directions. A method for tracking of the channel parameters with simultaneous transmission of the beams in the \ac{LOS} and \ac{NLOS} was introduced in \cite{ZhangC}. This method led to the reduced channel parameter accuracy compared to the link by link tracking \cite{VaV}. In addition, for the special cases that tracking information from the \ac{NLOS} links helps to improve the localization accuracy, it also considerably increases the number of required \ac{RF} chains. This is due to the fact that more than one beams are required to be transmitted in the direction of each first-order reflector to ensure a certain angular support during the tracking. Thus, a simplified position and rotation angle tracking system uses the information provided by the \ac{LOS} link \emph{only} with the tracking time\footnote{Note that due to the simultaneous transmission of beams $M_{\mathrm{ms}}\ll G$ in the uplink the tracking time is significantly reduced compared to the training time, i.e., $GT_{\mathrm{track}}\approx T_{\mathrm{train}}$.} $T_{\mathrm{track}}=2T_{s}$. Moreover, the position and rotation angle tracking accuracies are significantly reduced when the \ac{LOS} link is blocked \cite{Arash2x,DBLP:journals/corr/Abu-ShabanZASW17}. So, the \ac{LOS} link is used for tracking position and rotation angle and either \ac{LOS} detection methods or online blockage identification approaches can be adopted depending on the applications\footnote{As an example, blockage can be modeled based on 2-D knife edge diffraction \cite{QiW}. This allows the receiver to apply a transient change detection algorithm based on the mean change of a complex Gaussian distribution as the signal integrity metric for the \ac{LOS} blockage identification \cite{DaniEg}.} \cite{SchroederJ,DaniEg,Basseville,Hadjiliadis,PageE}. This allows to simply switch to the \ac{BS} that is in the \ac{LOS} thanks to the ultra dense networks.
\section{Heuristic Beamformer for Tracking}\label{Rob_Heur}
Previously, the robust beamformer design based on the minimization of the \ac{CRB} of the \ac{AOA}/\ac{AOD} for maximum angular spreads $\sigma^{\mathrm{max}}_{\theta_{0}}$ and $\sigma^{\mathrm{max}}_{\phi_{0}}$ was proposed in \cite{DBLP:journals/corr/GarciaWS17}. We propose a much simpler approach named as heuristic beamformers in beamspace. The proposed method does not require any optimization and relies on the angular support during the tracking phase. It is assumed that the beamformer $\mathbf{F}_{\mathrm{MS}}$ in the downlink is designed based on the maximum angular spread $\sigma_{\phi_{0}}^{\mathrm{max}}$. A similar process can be performed for the beamformer $\mathbf{F}_{\mathrm{BS}}$ in the uplink based on the maximum angular spread $\sigma_{\theta_{0}}^{\mathrm{max}}$. Moreover, the design of beamformers $\mathbf{F}^{[m-1]}_{\mathrm{MS}}$ in the uplink based on $\hat{\phi}_{0}^{[m-1]}$ and its uncertainty $[\mathbf{P}_{\boldsymbol{\psi}_{\mathrm{dl},0}}^{[m-1\vert m-1]}]_{1,1}$, and $\mathbf{F}^{[m-1]}_{\mathrm{BS}}$ in the downlink based on $\hat{\theta}_{0}^{[m-1]}$ and the uncertainty $[\mathbf{P}_{\boldsymbol{\psi}_{\mathrm{ul},0}}^{[m-1\vert m-1]}]_{2,2}$ follows a similar procedure. Finally, joint heuristic beamforming and channel estimation-tracking algorithm is proposed at the end of this section.
\subsection{Beamformer Design Problem}
We consider any time slots $[0, T_{\mathrm{ob}})$ where $T_{\mathrm{ob}}$ denotes the maximum duration for tracking phase, i.e., the observation time interval. The robust beamformer design within observation time interval $[0, T_{\mathrm{ob}})$ can be formulated as
\begin{IEEEeqnarray}{rCCl}\mathcal{R}_{\mathrm{B}}:\:\underset{\mathbf{F}_{\mathrm{MS}}}{\mbox{minimize}}& \:\underset{(\theta_{0},\phi_{0})\in \mathcal{R}_{\theta_{0}}\times \mathcal{R}_{\phi_{0}}}{\mbox{maximize}}&\:\mbox{maximize}\:\left\{\mathbb{E}\left[(\hat{\theta}_{0}-\theta_{0})^{2}\right],\mathbb{E}\left[(\hat{\phi}_{0}-\phi_{0})^{2}\right]\right\}\IEEEyesnumber\IEEEyessubnumber\label{Rob_Prob2}\\
\mbox{subject to} &\qquad & \mathbb{E}\left[\hat{\boldsymbol{\eta}}_{0}\right]=\boldsymbol{\eta}_{0},\IEEEyessubnumber\label{Rob_Prob3}
\end{IEEEeqnarray}
where $\mathcal{R}_{\theta_{0}}$ and $\mathcal{R}_{\phi_{0}}$ denote the ranges of $\theta_{0}$ and $\phi_{0}$ for the maximum angular spreads $\sigma^{\mathrm{max}}_{\theta_{0}}$ and $\sigma^{\mathrm{max}}_{\phi_{0}}$, respectively, and the constraint \eqref{Rob_Prob3} guarantees the estimation of the channel parameters for the \ac{LOS} path $\hat{\boldsymbol{\eta}}_{0}$ to be unbiased. The expectations and unbiased constraints in $\mathcal{R}_{\mathrm{B}}$ can be replaced using the equivalent terms from the \ac{FIM}. Consequently, the robust beamformer $\mathbf{F}_{\mathrm{MS}}=\mathbf{F}_{\mathrm{RF},\mathrm{MS}}\mathbf{F}_{\mathrm{BB},\mathrm{MS}}$ is designed for maximum angular spreads $\sigma^{\mathrm{max}}_{\theta_{0}}$ and $\sigma^{\mathrm{max}}_{\phi_{0}}$ within the observation time $T_{\mathrm{ob}}$ \cite{DBLP:journals/corr/GarciaWS17}. 
\subsection{Heuristic Beamformer}
To provide angular support of the robust beamformer within the observation time $T_{\mathrm{ob}}$ for the \ac{LOS}\cite{DBLP:journals/corr/GarciaWS17}, one needs to use the estimated \ac{AOA} in the downlink from the training phase. Then, the heuristic beamformer for even and odd number of transmitted beams $M_{\mathrm{MS}}$ are obtained as\footnote{The transmit beamformers in the downlink and uplink are obtained similarly using the sequential transmission in case of limited number of \ac{RF} chains, or transmission in parallel otherwise.}
\begin{equation}\label{Heuristic1}
\setlength{\nulldelimiterspace}{0pt}
\mathbf{F}_{\mathrm{MS},\mathrm{heu}}=\left\{\begin{IEEEeqnarraybox}[\relax][c]{l's}
\mathbf{A}_{\mathrm{odd}}\left(\hat{\phi}_{0},\left\{\hat{\phi}_{0}\pm i\Delta\phi_{3\mathrm{dB}}\right\}\right),& $M_{\mathrm{MS}}= \mathrm{odd}$ and $i=1,\ldots, \frac{M_{\mathrm{MS}}-1}{2}$\\
\mathbf{A}_{\mathrm{even}}\left(\hat{\phi}_{0}\pm 0.5\Delta\phi_{3\mathrm{dB}},\left\{\hat{\phi}_{0}\pm i\Delta\phi_{3\mathrm{dB}}\right\}\right),& $M_{\mathrm{MS}}= \mathrm{even}$ and $i=2,\ldots, \frac{M_{\mathrm{MS}}}{2}$
\end{IEEEeqnarraybox}\right.
\end{equation}
where $\mathbf{A}_{\mathrm{odd}}\left(\hat{\phi}_{0},\left\{\hat{\phi}_{0}\pm i\Delta\phi_{3\mathrm{dB}}\right\}\right)$ and $\mathbf{A}_{\mathrm{even}}\left(\hat{\phi}_{0}\pm 0.5\Delta\phi_{3\mathrm{dB}},\left\{\hat{\phi}_{0}\pm i\Delta\phi_{3\mathrm{dB}}\right\}\right)$ denote the $N_{\mathrm{MS}}\times M_{\mathrm{MS}}$ matrices with 3 dB beam overlap, the columns formed as shifted versions of $\mathbf{a}_{\mathrm{MS}}(\hat{\phi}_{0})$, and even and odd number of beams, respectively, the value of $M_{\mathrm{MS}}$ should approximately satisfy $M_{\mathrm{MS}}\geq\lceil\sigma^{\mathrm{max}}_{\phi_{0}}N_{\mathrm{MS}}/2\rceil$ for the symmetric coverage of the maximum angular spread of $\sigma^{\mathrm{max}}_{\phi_{0}}$ centered around $\hat{\phi}_{0}$, and $\Delta\phi_{3\mathrm{dB}}$ denotes the phase shift to provide the 3 dB beam overlap for the \ac{ULA} \cite{OrfandisJ}.
As an example, for the maximum angular spread of $\sigma^{\mathrm{max}}_{\phi_{0}}\:[\mathrm{deg}]=20$ centered around $\hat{\phi}_{0}$ and $N_{\mathrm{MS}}=32$, one needs at least $M_{\mathrm{MS}}=6$ beams. In principle, heuristic beamformer is obtained by forming $M_{\mathrm{MS}}$ beams centered around the estimated \ac{AOA} in the downlink obtained from the training phase. Since, the 3 dB beam overlap is applied the best spatial coverage is obtained in the tracking phase. 
\begin{figure}
\centering
\psfrag{optimal robust}{\tiny Optimal robust}
\psfrag{optimal heuristic}{\tiny Optimal heuristic}
\psfrag{hybrid robust precoder with 1-bit adaptive network}{\tiny Hybrid robust precoder with 1-bit adaptive network}
\psfrag{hybrid heuristic precoder with 1-bit adaptive network}{\tiny Hybrid heuristic precoder with 1-bit adaptive network}
\psfrag{CRB(theta) [deg]}{\footnotesize $\mathrm{rsCRB}(\hat{\phi}_{0})\:[\mathrm{deg}]$}
\psfrag{SNR [dB]}{\footnotesize $\mathrm{SNR}$ (in dB)}
\includegraphics[width=0.45\columnwidth]{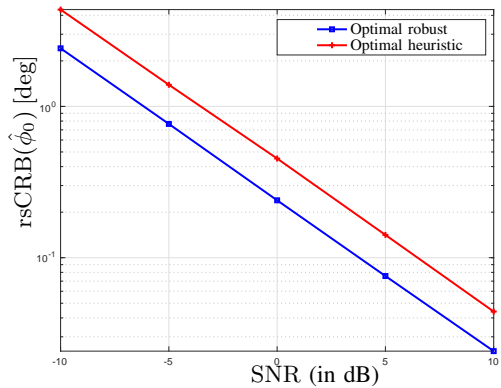}
\caption{Comparison between the performance of robust and heuristic beamformers versus \ac{SNR}.}
\label{boundcomp}
\end{figure}
\subsection{Numerical Comparison}
Unlike the method in \cite{DBLP:journals/corr/GarciaWS17} for the robust beamformer design, the proposed heuristic solution does not require to solve a complex optimization problem.
The performance in terms of \ac{rsCRB} for the maximum angular spread of $\sigma^{\mathrm{max}}_{\phi_{0}}\:[\mathrm{deg}]=20$ is compared in Fig. \ref{boundcomp}. It is observed that the heuristic design leads to around $50\%$ relative increase of $\mathrm{rsCRB}(\hat{\phi}_{0})\:[\mathrm{deg}]$ as the price for reducing the complexity compared to the robust counterparts. 
\begin{algorithm}[h!]
\SetKwInOut{Input}{Input}
\SetKwInOut{Output}{Output}
 \Input{Set $m=1$, and $T_{\mathrm{ob}}$.}
 \Output{Tracked position $\hat{\mathbf{p}}^{[m]}$ and rotation angle $\hat{\alpha}^{[m]}$ within $T_{\mathrm{ob}}$ with the corresponding uncertainties in \eqref{geom_track4} and \eqref{geom_track5}, respectively.}
\Repeat{the next observation time $T_{\mathrm{ob}}$}{
Compute $\hat{\tau}_{0}^{[m-1]}$ and $\hat{\theta}_{0}^{[m-1]}$ in the \ac{BS}, $\hat{\phi}_{0}^{[m-1]}$ in the \ac{MS}, and the corresponding values of $\hat{\mathbf{p}}^{[m-1]}$ and $\hat{\alpha}^{[m-1]}$ in the training\;
Hybrid implementation of the received beamformer in the downlink \eqref{Heuristic1} and similarly received beamformer in the uplink\;
\While{$mT_{B}\leq T_{\mathrm{ob}}$}{
Design $\mathbf{F}^{[m-1]}_{\mathrm{MS}}$ in the uplink based on $\hat{\phi}_{0}^{[m-1]}$ and $[\mathbf{P}_{\boldsymbol{\psi}_{\mathrm{dl},0}}^{[m-1\vert m-1]}]_{1,1}$\;
Compute $\hat{\tau}_{0}^{[m]}$, $\hat{\theta}_{0}^{[m]}$, and $\hat{\mathbf{p}}^{[m]}$ in the BS\;
Design $\mathbf{F}^{[m-1]}_{\mathrm{BS}}$ in the downlink based on $\hat{\theta}_{0}^{[m-1]}$ and $[\mathbf{P}_{\boldsymbol{\psi}_{\mathrm{ul},0}}^{[m-1\vert m-1]}]_{2,2}$\;
Compute $\hat{\phi}_{0}^{[m]}$ in the \ac{MS}\;
Using $\hat{\mathbf{p}}^{[m]}$ and $\hat{\phi}_{0}^{[m]}$, compute $\hat{\alpha}^{[m]}$ in the \ac{MS}\;
Set $m=m+1$\;
}
}
\caption{Heuristic Beamforming and Position and Orientation Training-Tracking\label{beam_ch_track}}
\end{algorithm}
\subsection{Hybrid Design}
Similar to \cite{AyachOE}, the hybrid design solution for the proposed heuristic beamformer in \eqref{Heuristic1} is obtained using sparsity constrained matrix reconstruction problem that can be solved using the \ac{OMP} type algorithms. Consequently, the baseband precoder is obtained by $M_{\mathrm{MS}}$ nonzero rows of $\mathbf{F}_{\mathrm{BB},\mathrm{MS}}$ with $M_{\mathrm{MS}}$ columns and $\mathbf{F}_{\mathrm{RF},\mathrm{MS}}$ is given by the corresponding columns of $\mathbf{U}_{\mathrm{MS}}$ for the ideal beam selection, and mapped to the closest 1-bit beamformer with $\{+1,-1\}$ for 1-bit beam selection\cite{Donnodd}.
\subsection{Heuristic Beamforming and Position and Orientation Training-Tracking}
The heuristic beamforming and position and orientation training-tracking algorithm is summarized in the Algorithm \ref{beam_ch_track}. In step 2, the estimated channel parameters $\hat{\boldsymbol{\eta}}_{0}^{[m-1]}=[\hat{\tau}_{0}^{[m-1]},\hat{\theta}_{0}^{[m-1]},\hat{\phi}_{0}^{[m-1]}]^{\mathrm{T}}$ and corresponding values of the position $\hat{\mathbf{p}}^{[m-1]}$ and rotation angle $\hat{\alpha}^{[m-1]}$ are obtained\footnote{For $m=1$, these values are obtained from the training phase, and for $m>1$ they are obtained from the tracking phase.}. In step 3, the heuristic hybrid beamformer with 1-bit phase shifters is applied for the tracking phase for the receiver in the uplink/downlink. In step 5, the uplink beamformer $\mathbf{F}^{[m-1]}_{\mathrm{MS}}$ is designed based on the previous downlink \ac{AOA} and the corresponding uncertainty. In step 6, the \ac{BS} tracks the \ac{TOA}-\ac{AOA} and the location of the \ac{MS} in the uplink. In step 7, the downlink beamformer $\mathbf{F}^{[m-1]}_{\mathrm{BS}}$ is designed based on the previous uplink \ac{AOA} and the corresponding uncertainty. In step 8, the \ac{MS} tracks the \ac{AOA} in the downlink. In step 9, the rotation angle is obtained using the location information that is fed back to the \ac{MS} and the \ac{AOA} in the downlink, and the block index is updated in step 10 untill $mT_{B}\leq T_{\mathrm{ob}}$. Finally, the steps 2-11 are repeated for the next observation time $T_{\mathrm{ob}}$.
\section{Simulation Results}\label{Sim_Res}
In this section, the performance of the proposed algorithms for different parameters is investigated.
\subsection{Simulation Setup}\label{setup}
We consider a scenario representative of indoor localization based on METIS Madrid grid model\cite{Maltsevx}. We employ the ray tracing simulation tool in order to model the propagation of signals in the uplink and downlink for channel training and tracking \cite{PyLayer}. 
We set $f_c\:[\mathrm{GHz}]= 60$, $B\:[\mathrm{MHz}]=200$, $c\:[\mathrm{m}/\mathrm{ns}]=0.299792$, and $N=40$. The number of transmit and receive antennas are set to $N_{\mathrm{BS}}=32$ and $N_{\mathrm{MS}}=32$, respectively, unless otherwise stated. The number of simultaneous beams for the training are set to $M_{\mathrm{BS}}=1$ and $M_{\mathrm{MS}}=1$ for the downlink and the uplink, respectively. The number of sequentially transmitted signals for training in the downlink and the uplink satisfies $G\succeq (\hat{K}+1)\log (N_{\mathrm{MS}/\mathrm{BS}}/(\hat{K}+1))$. Unless otherwise stated, the required number of elements for the \ac{SD}-based training is set to $V=3$. Finally, the received \ac{SNR} in the downlink is defined as
\begin{equation}\label{SNRdef}
\mathrm{SNR}=\frac{\mathbb{E}\left[\left\Vert\mathrm{blkdiag}\left(\{\mathbf{\Omega}_{\mathrm{MS}}[n]\}_{n=0}^{N-1}\right)\mathrm{vec}\left(\{\check{\mathbf{h}}_{\mathrm{MS}}[n]\}_{n=0}^{N-1}\right)\right\Vert^{2}_{2}\right]}{\mathbb{E}\left[\left\Vert\mathrm{vec}\left(\{\check{\mathbf{n}}_{\mathrm{MS}}[n]\}_{n=0}^{N-1}\right)\right\Vert^{2}_{2}\right]},
\end{equation}
where $\mathrm{blkdiag}\left(.\right)$ creates a block diagonal matrix from its arguments and $\mathrm{vec}\left(.\right)$ creates a tall column vector from its arguments. Similarly, the received \ac{SNR} in the uplink is defined by replacing the subscript MS by BS. During the tracking phase and unless stated otherwise, the \ac{MS} moves with the velocity on the order of $4\:\mathrm{m}/\mathrm{s}$ that is in the high range of velocity for indoor localization. Constant angular rates for the \ac{UL-AOA} and \ac{DL-AOA} are on the order of $0.4529\:\mathrm{deg}/T_{B}$ and $0.2265\:\mathrm{deg}/T_{B}$, respectively. The block duration and the observation time are on the order of $T_{B}\:[\mathrm{ms}]=10$ and $T_{\mathrm{ob}}\:[\mathrm{s}]=0.6$, respectively. The maximum angular spreads are set to $\sigma^{\mathrm{max}}_{\theta_{0}}\:[\mathrm{deg}]=\sigma^{\mathrm{max}}_{\phi_{0}}\:[\mathrm{deg}]=20$ centered around $\hat{\theta}_{0}$ and $\hat{\phi}_{0}$ within the simulations. During the tracking, $M_{\mathrm{MS}}$ in the downlink and $M_{\mathrm{BS}}$ in the uplink are set to guarantee the aforementioned maximum angular supports, e.g., $M_{\mathrm{MS}}=7$ in the downlink and $M_{\mathrm{BS}}=7$ in the uplink for $N_{\mathrm{BS}}=N_{\mathrm{MS}}=32$. Unless otherwise stated, the power of the process noise for the continuous-time state model is set to $\mathbf{Q}_{c}=\mathrm{diag}\{\sigma^{2}_{\tau_{0}},\sigma^{2}_{\theta_{0}},\sigma^{2}_{\phi_{0}}\}$ with the standard deviations $\sigma_{\tau_{0}}\:[\mathrm{ns}]=0.5$ and $\sigma_{\theta_{0}}\:[\mathrm{deg}]=\sigma_{\phi_{0}}\:[\mathrm{deg}]=5$, and $\mathbf{\Phi}$ and $\mathbf{Q}^{[m]}_{0}$ are obtained by numerical discretization\cite{DBLP:journals/twc/KoivistoCWHTLKV17}.

The performance of the maximum \ac{RMSE} within the observation time $T_{\mathrm{ob}}$ for the estimation and tracking algorithms was assessed from $100$ Monte Carlo realizations. The false alarm probability was set to ${P}_{\mathrm{fa}}=10^{-3}$ to determine the threshold $\delta$ in \eqref{Threshold_express}. 
\begin{figure}   
\psfrag{Proposed extended SD-based estimation, V=3}{\tiny Proposed extended SD-based estimation, $V=3$}
\psfrag{Proposed extended SD-based estimation, V=5}{\tiny Proposed extended SD-based estimation, $V=5$}
\psfrag{Conventional CS-based estimation, V=3}{\tiny Conventional CS-based estimation, $V=3$}
\psfrag{Conventional CS-based estimation, V=5}{\tiny Conventional CS-based estimation, $V=5$}
\psfrag{dr}{\footnotesize $\delta_{r}$}
\psfrag{SNR [dB]}{\footnotesize $\mathrm{SNR}$ (in dB)}
\centering
\includegraphics[width=0.5\columnwidth]{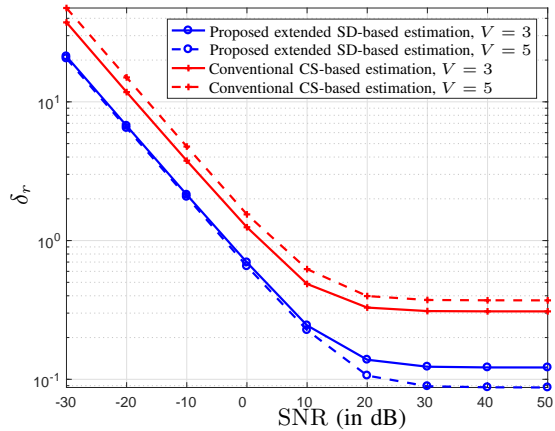}
  \caption{Residual error for the proposed extended SD-based and conventional \ac{CS}-based estimations versus \ac{SNR} for $V=\{3,5\}$.}
  \label{V_Effects}
\end{figure}
\subsection{Results and Discussion}\label{results}
\subsubsection*{Performance Comparison with Respect to the Adopted CS-Based Estimation for Lens Array}
Fig. \ref{V_Effects} compares the accuracy in terms of the residual error $\delta_{r}$ of the proposed extended \ac{SD}-based estimation algorithm \ref{algor0_det} with the adopted \ac{CS}-based estimation for lens array with different values of \ac{SNR} and the number of elements $V=\{3,5\}$. It is observed that the residual error in \ac{CS}-based estimation is approximately twice the corresponding values of the proposed \ac{SD}-based algorithm. This effect is most pronounced in the low SNR regime, i.e., $\mathrm{SNR}<0$ dB, as the gap between the residual error of the proposed extended \ac{SD}-based algorithm and the \ac{CS}-based estimation is increased. Moreover, by increasing the number of elements $V$ the error in the \ac{CS}-based estimation is increased compared to the proposed extended \ac{SD}-based method. This is due to the reduction of detection probability of the support of the sparse channel coefficient 

\begin{figure}
 \centering
\psfrag{Heuristic tracking with refined training}{\tiny Tracking with refined training \& heuristic beamforming}
\psfrag{Heuristic tracking with non-refined training}{\tiny Tracking with non-refined training \& heuristic beamforming}
\psfrag{PEB}{\tiny PEB}
\psfrag{REB}{\tiny REB}
\psfrag{SNR [dB]}[c][c]{\footnotesize $\mathrm{SNR}$ (in dB)}
\psfrag{RMSE(p) [m]}[c][c]{\footnotesize $\mathrm{RMSE}_{\mathrm{max}}(\hat{\mathbf{p}}^{[m]})$ [m]}
\psfrag{RMSE(alpha) [deg]}[c][c]{\footnotesize $\mathrm{RMSE}_{\mathrm{max}}(\hat{\alpha}^{[m]})\:[\mathrm{deg}]$}
\includegraphics[width=0.55\linewidth]{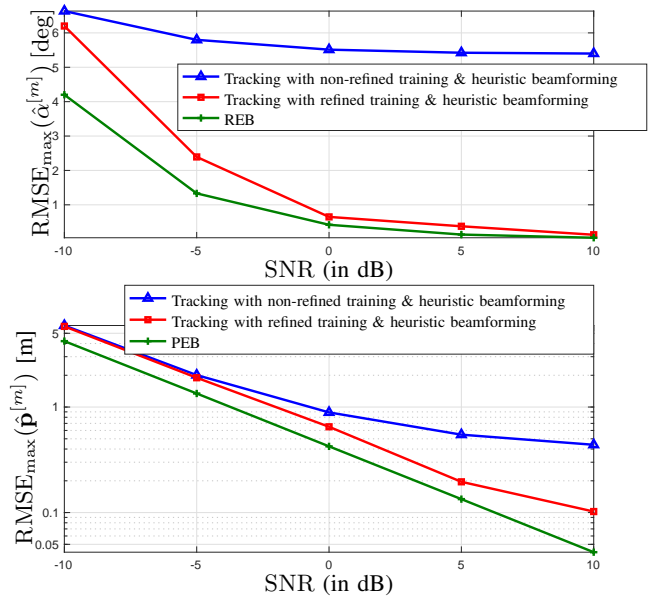}
 \caption{The maximum \ac{RMSE} of the \ac{MS} (top) rotation angle $\hat{\alpha}^{[m]}$, the corresponding REB, (bottom) position $\hat{\mathbf{p}}^{[m]}$, and the corresponding PEB after training with/without refinement and tracking within the observation time $\forall m\in\mathcal{M}_{\mathrm{ob}}$ for $\mathrm{SNR}\:[\mathrm{dB}]=\{-10, -5, 0, 5, 10\}$ and $N_{\mathrm{MS}}=N_{\mathrm{BS}}=32$.}\label{fig_tracking1} 
\end{figure}
\subsubsection*{Performance Comparison with Respect to \ac{SNR}}
Fig. \ref{fig_tracking1} shows the performance of the proposed algorithms for training and tracking for $\mathrm{SNR}\:[\mathrm{dB}]=\{-10, -5, 0, 5, 10\}$ and $N_{\mathrm{MS}}=N_{\mathrm{BS}}=32$. The performance of the proposed training and tracking algorithms with heuristic beamformers in the tracking phase with the refinement stated in Section III.C and without refinement in the training phase is compared to the corresponding values of the \ac{REB} and the \ac{PEB} within the observation time, i.e., $\forall m\in\mathcal{M}_{\mathrm{ob}}$, for different values of \ac{SNR} \cite{Arash2x,Arash}. In Fig. \ref{fig_tracking1} top plot, it is observed that after $\mathrm{SNR}\:[\mathrm{dB}]=0$ the maximum value of \ac{RMSE} of the rotation angle within the observation time saturates to around 5 degrees for the non-refined training and increasing the \ac{SNR} does not improve the rotation angle estimation accuracy anymore. This is due to the limited number of antenna elements in the \ac{BS} and the \ac{MS}. However, using the refinement in the training phase it is possible to approach the corresponding values of the \ac{REB} without significant saturation. Moreover, the \ac{RMSE} of the \ac{MS} position approaches the \ac{PEB} for $\mathrm{SNR}\:[\mathrm{dB}]=0$ within the observation time. The gap between the \ac{RMSE} of \ac{MS} position and the \ac{PEB} within the observation time is mainly due to the limited number of antenna elements in the estimation of \ac{AOD} as one of the key factors for localization that can be improved by applying the refinement in the training phase as shown in the bottom plot. 
\begin{figure}   
\psfrag{RMSE(p) [m]}[c][c]{\footnotesize $\mathrm{RMSE}_{\mathrm{max}}(\hat{\mathbf{p}}^{[m]})\:[\mathrm{m}]$}
\psfrag{RMSE(alpha) [deg]}[c][c]{\footnotesize $\mathrm{RMSE}_{\mathrm{max}}(\hat{\alpha}^{[m]})\:[\mathrm{deg}]$}
\psfrag{Nbs=Nms}[c][c]{\footnotesize $N_{\mathrm{MS}}=N_{\mathrm{BS}}$}
\psfrag{Heuristic tracking with non-refined training}{\tiny Tracking with non-refined training \& heuristic beamforming}
\psfrag{Heuristic tracking with refined training}{\tiny Tracking with refined training \& heuristic beamforming}
\psfrag{REB}[][c]{\tiny REB}
\psfrag{PEB}[][c]{\tiny PEB}
\centering
\label{Nt_Effectsa}\includegraphics[width=0.55\columnwidth]{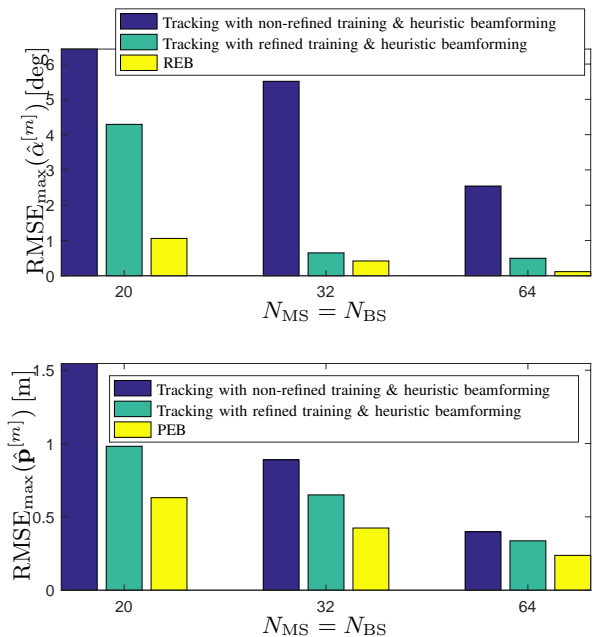}
  \caption{The maximum \ac{RMSE} of the \ac{MS} (top) rotation angle $\hat{\alpha}[m]$ and the corresponding REB, (bottom) position $\hat{\mathbf{p}}^{[m]}$ and the corresponding PEB, after training with/without refinement and tracking within the observation time $\forall m\in\mathcal{M}_{\mathrm{ob}}$ for $N_{\mathrm{MS}}=N_{\mathrm{BS}}=\{20, 32, 64\}$ and $\mathrm{SNR}\:[\mathrm{dB}]=0$.}
  \label{Nt_Effects}
\end{figure}
\subsubsection*{Performance Comparison with Respect to the Number of Antenna Elements}
Fig. \ref{Nt_Effects} shows the performance of the training with/without refinement and tracking algorithms for the rotation angle and position of the \ac{MS} for $N_{\mathrm{MS}}=N_{\mathrm{BS}}=\{20, 32, 64\}$ at $\mathrm{SNR}\:[\mathrm{dB}]=0$. The performance of the proposed training and tracking algorithms with heuristic beamformer is compared to the corresponding values of the \ac{REB} and \ac{PEB} within the observation time for different number of antenna elements \cite{Arash2x,Arash}. In Fig. \ref{Nt_Effects}, the accuracies of the rotation angle and position after training without refinement and tracking in the observation time $T_{\mathrm{ob}}$ are mainly limited by the angular resolution that depends on the number of antenna elements, e.g., for $N_{\mathrm{MS}}=N_{\mathrm{BS}}=32$ it is possible to achieve the rotation angle accuracy of around 5 degrees. This is the main reason for the gap between the proposed method without refinement in the training phase and the corresponding values of the \ac{REB} and PEB within the observation time. In principle, one can apply the angular refinement procedure in the training phase to obtain closer values to the corresponding \ac{REB} and \ac{PEB} as shown with the green bars. 
\subsubsection*{Performance Comparison with Respect to the Power of the Process Noise}
\begin{figure}   
\psfrag{RMSE(p) [m]}[c][c]{\footnotesize $\mathrm{RMSE}(\hat{\mathbf{p}}^{[m]})\:[\mathrm{m}]$}
\psfrag{RMSE(alpha) [deg]}[c][c]{\footnotesize $\mathrm{RMSE}(\hat{\alpha}^{[m]})\:[\mathrm{deg}]$}
\psfrag{siga=sigd=2 deg}{\tiny $\sigma_{\theta_{0}}=\sigma_{\phi_{0}}=2$ deg}
\psfrag{siga=sigd=5 deg}{\tiny $\sigma_{\theta_{0}}=\sigma_{\phi_{0}}=5$ deg}
\psfrag{siga=sigd=12.5 deg}{\tiny $\sigma_{\theta_{0}}=\sigma_{\phi_{0}}=12.5$ deg}
\psfrag{Block index}{\footnotesize Block index $m$}
\centering
\label{Block_effects}\includegraphics[width=0.55\columnwidth]{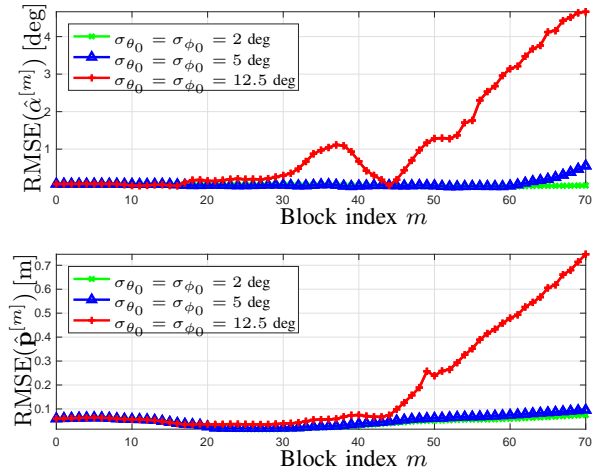}
  \caption{The \ac{RMSE} of the \ac{MS} (top) rotation angle $\hat{\alpha}[m]$ and (bottom) position $\hat{\mathbf{p}}^{[m]})$ with the given rate of changes, and the standard deviation of the process noise with the diagonal components of $\sigma_{\tau_{0}}\:[\mathrm{ns}]=0.5$ and $\sigma_{\theta_{0}}\:[\mathrm{deg}]=\sigma_{\phi_{0}}\:[\mathrm{deg}]=\{2,5,12.5\}$ after training with refinement for the first $70$ block indices with $N_{\mathrm{MS}}=N_{\mathrm{BS}}=32$ and $\mathrm{SNR}\:[\mathrm{dB}]=10$.}
  \label{Block_effectsx}
\end{figure}

Fig. \ref{Block_effectsx} shows the performance of the training with refinement and tracking algorithms using heuristic design with respect to the power of the process noise for the aforementioned rate of changes. It is worth noting that the change of delay only affects tracking of the position and does not influence the rotation angle tracking. Consequently, the effects of different values of noise standard deviation of \ac{UL-AOA} and \ac{DL-AOA}, i.e., $\sigma_{\theta_{0}}$ and $\sigma_{\phi_{0}}$, are analyzed for a given value of $\sigma_{\tau_{0}}$. The components of the standard deviation of the process noise for the continuous-time state model $\mathbf{Q}_{c}$ are set to $\sigma_{\tau_{0}}\:[\mathrm{ns}]=0.5$ and $\sigma_{\theta_{0}}\:[\mathrm{deg}]=\sigma_{\phi_{0}}\:[\mathrm{deg}]=\{2,5,12.5\}$. For $m>20$, it is observed that the \ac{RMSE} of position and rotation angle gradually increases versus the block index by increasing the standard deviations of \ac{UL-AOA} and \ac{DL-AOA} to $\sigma_{\theta_{0}}\:[\mathrm{deg}]=\sigma_{\phi_{0}}\:[\mathrm{deg}]=12.5$ compared to the estimated values, i.e., block index zero. This is mainly due to the limited angular support of the received beamformers designed based on the maximum angular spreads $\sigma^{\mathrm{max}}_{\theta_{0}}\:[\mathrm{deg}]=\sigma^{\mathrm{max}}_{\phi_{0}}\:[\mathrm{deg}]=20$. On the other hand, for low to average values of the standard deviations of \ac{UL-AOA} and \ac{DL-AOA} on the order of $\sigma_{\theta_{0}}\:[\mathrm{deg}]=\sigma_{\phi_{0}}\:[\mathrm{deg}]=\{2,5\}$ the \ac{RMSE} of position and rotation angle are close to the values obtained from the training during $1\leq m<60$, i.e., block index zero. The main reason is that the user is mostly moving within the the angular support of the received beamformers. For $m\geq 60$ and $\sigma_{\theta_{0}}\:[\mathrm{deg}]=\sigma_{\phi_{0}}\:[\mathrm{deg}]=5$, the user starts to move out of the angular support that leads to increasing the \ac{RMSE}. 
\subsubsection*{Performance Comparison with the Method in \cite{VaV}}
\begin{figure}   
\psfrag{py [m]}[c][c]{\footnotesize $\hat{p}^{[m]}_{y}$ [m]}
\psfrag{px [m]}[c][c]{\footnotesize $\hat{p}^{[m]}_{x}$ [m]}
\psfrag{Proposed}{\tiny Proposed}
\psfrag{Method in Heath}{\tiny Method in \cite{VaV}}
\psfrag{Trajectory}{\tiny Trajectory}
\psfrag{alpha [deg]}{\footnotesize $\hat{\alpha}^{[m]}\:[\mathrm{deg}]$}
\psfrag{Block index}{\footnotesize Block index $m$}
\psfrag{siga=sigd=2 deg}{\tiny $\sigma_{\theta_{0}}=\sigma_{\phi_{0}}=2$ deg}
\psfrag{siga=sigd=5 deg}{\tiny $\sigma_{\theta_{0}}=\sigma_{\phi_{0}}=5$ deg}
\centering
\label{Block_effects}\includegraphics[width=0.63\columnwidth]{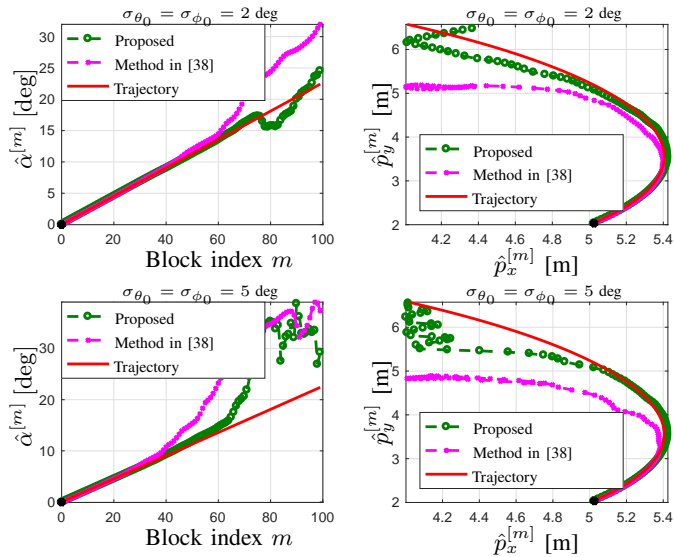}
  \caption{The trajectory of the \ac{MS} rotation angle $\hat{\alpha}[m]$, and position $\hat{\mathbf{p}}^{[m]})$ with $N_{\mathrm{MS}}=N_{\mathrm{BS}}=32$ and $\mathrm{SNR}\:[\mathrm{dB}]=10$ compared to the method in \cite{VaV} for (top) $\sigma_{\theta_{0}}\:[\mathrm{deg}]=\sigma_{\phi_{0}}\:[\mathrm{deg}]=2$, and (bottom) $\sigma_{\theta_{0}}\:[\mathrm{deg}]=\sigma_{\phi_{0}}\:[\mathrm{deg}]=5$. The black stars show the estimated position and rotation angle from the training.}
  \label{complitt}
\end{figure}

For the sake of comparison, the method in \cite{VaV} is adopted to the more complete state model including the rate of change terms in \eqref{track2} with lens antenna arrays in the transmitter and receiver for $\sigma_{\tau_{0}}\:[\mathrm{ns}]=0.5$ and $\sigma_{\theta_{0}}\:[\mathrm{deg}]=\sigma_{\phi_{0}}\:[\mathrm{deg}]=\{2,5\}$. In Fig. \ref{complitt}, it is observed that the proposed method provides a more robust performance for position and rotation angle tracking within the observation time. This is mainly due to the fact that the received beamformers in the proposed method are designed based on the maximum angular spreads in the uplink and downlink. For $\sigma_{\theta_{0}}\:[\mathrm{deg}]=\sigma_{\phi_{0}}\:[\mathrm{deg}]=2$ and $m>70$, the user is out of the angular support in the proposed method and starts diverging from the trajectory. By increasing the angular spreads to  $\sigma_{\theta_{0}}\:[\mathrm{deg}]=\sigma_{\phi_{0}}\:[\mathrm{deg}]=5$, diverging from the trajectory occurs slightly faster at $m>60$ in the proposed method. In \cite{VaV}, the design is not based on the maximum angular spreads. This leads to beam misalignment and non-robust performance compared to the proposed method and consequently faster divergence from the trajectory. 

\section{Conclusion}\label{SEC:Conclusion}
We have studied the determination of a receiver position and orientation using a single transmitter in a \ac{mm-wave} \ac{MIMO} system with lens antenna arrays. Our study includes novel solutions for training the \ac{mm-wave} lens \ac{MIMO}     channel, tracking position and orientation, and heuristic beamformer design. We have proposed a novel method for training of the \ac{mm-wave} lens \ac{MIMO} channel with the reduced training time. A simplified approach named heuristic beamformer design was proposed for position and orientation tracking. The proposed heuristic beamformer does not require to solve complex optimization problems. Using the proposed heuristic design and tracking algorithm, a joint beamformer design and position and orientation tracking method was proposed. Through simulation studies, we have demonstrated the efficiency of the proposed algorithms, and have shown that the proposed training and joint beamformer design and tracking methods provide practical solutions for updating the location information of the user in dynamic conditions. In particular, the performance of the proposed position and rotation angle tracking is close to the estimated values during the observation time without re-estimating channel parameters for different blocks with much reduced complexity.
\section{Complexity Analysis}\label{app_A}
We analyze the complexity of different stages of the proposed training phase.\begin{itemize}\item Algorithm \ref{algor0_det}: The complexity in performing \eqref{Training15} is on the order of $O(N_{\mathrm{MS}}GN_{\mathrm{sub}})$ where $N_{\mathrm{sub}}$ denotes the few subcarriers sufficient to detect the dominant path. The \ac{LS} solution in \eqref{Training18} can be computed with the complexity on the order of $O(GNV^{2})$ for all the subcarriers. The complexity of the matrix $\bar{\mathbf{F}}_{\mathrm{MS},x}[n]$ of size $G\times N_{\mathrm{MS}}$ multiplied by the vector $\hat{\check{\mathbf{h}}}_{\mathrm{MS},t}[n]\big\vert_{\mathrm{supp}\left(\boldsymbol{\chi}_{\mathrm{MS},t}\right)}$ of size $N_{\mathrm{MS}}\times 1$ in \eqref{Training19} is on the order of $O(GNN_{\mathrm{MS}})$ for all the subcarriers. Finally, the complexity in computing $\tau_{k}$ is on the order of $O(ND_{o}\hat{K})$ where $D_{o}$ denotes the number of delay grid points, and computing $\tilde{h}_{k}$ requires $O(N\hat{K})$ operations. Consequently, the maximum complexity of the Algorithm \ref{algor0_det} is dominated by the term $\hat{K}\times O(GNN_{\mathrm{MS}})$.
\item \ac{AOA} estimation in the uplink: Similarly, the total complexity for the estimation of \ac{AOA} in the uplink is dominated by the term $\hat{K}\times O(N_{\mathrm{BS}}GN_{\mathrm{sub}})$. 
\end{itemize}
Following the above discussion\footnote{To reduce the computational complexity in the \ac{MS}, it is possible to start the training phase in the uplink for joint estimation of the \ac{AOA} and \ac{TOA} at the \ac{BS}. Due to channel reciprocity, the same algorithms in the training phase can be easily adopted in this case.}, the overall complexity for channel training is approximately on the order of $\hat{K}\times\left(O(GNN_{\mathrm{MS}})+O(N_{\mathrm{BS}}GN_{\mathrm{sub}})\right)$. 
\bibliographystyle{IEEEtran}
\bibliography{References_Manuscript}
\end{document}